\documentclass[sigconf,10pt]{acmart}
\usepackage{xspace}

\newcommand{\sysname}{\textsc{HeteroVDiff}\xspace}

\newcommand{\eg}{e.g.,\xspace}
\AtBeginDocument{%
  }

\setcopyright{acmlicensed}
\copyrightyear{2018}
\acmYear{2018}
\acmDOI{XXXXXXX.XXXXXXX}
\acmConference[Conference acronym 'XX]{Make sure to enter the correct
  conference title from your rights confirmation email}{June 03--05,
  2018}{Woodstock, NY}
\acmISBN{978-1-4503-XXXX-X/2018/06}

\begin{document}

\title[Pipelining Video Diffusion with Shared Weights]{HeteroVDiff: Pipelining Video Diffusion across Mobile NPU and GPU with Shared Weights}

\author{Zheyu Shen, Ziyao Wang, Yexiao He, Guoheng Sun, Yuning Zhang, Yiting Wang, Wanghao Ye, Ang Li}
\affiliation{%
  \institution{University of Maryland, College Park}
  \city{College Park}
  \state{MD}
  \country{USA}}
\email{{zyshen,ziyaow,yexiaohe,ghsun,yuning,ywang144,wy891,angliece}@umd.edu}

\renewcommand{\shortauthors}{Shen et al.}

\begin{abstract}
Video generation is moving from the cloud to phones for privacy and offline use, and the applications often ask for several clips at a time. 
So video generation on phones is a throughput problem before it is a latency problem.
Contemporary mobile SoCs carry both a GPU and a neural processing unit (NPU) on shared DRAM.
Yet on-device video diffusion systems run the denoiser on one of them.
The other engine either sits idle or only decodes the finished latent afterwards.
The heterogeneous designs developed for mobile LLMs do not carry over, because they synchronise the two engines at every operator and partition weight tensors between the engines rather than sharing one stored copy.
In this paper we first measure the platform limits that decide what can run on this class of SoC.
They include how large one buffer may be, whether both runtimes can map the same buffer, and how much activation memory the NPU toolchain lets one graph hold.
Drawing on these measurements, we store the base weights once in INT8 in the layout the NPU reads natively.
The GPU uses the same bytes directly as integer dot-product operands, so neither runtime rearranges the weights on load.
We then flatten the denoising trajectory into its full sequence of block executions and cut it once at a block boundary, so the GPU runs the prefix and the NPU the suffix.
We place the cut from measured rather than modelled rates.
Those rates include the decoder that only the NPU can run.
They are taken while both engines run together, because each engine's speed changes when the other is active.
By employing these techniques we present \sysname, which serves a multi-expert video diffusion model across the NPU and the GPU using 48\% less weight memory than giving each engine its own copy.
Generating videos back to back on a Galaxy~S25 with decoding inside the pipeline, \sysname sustains $1.35\times$ the stream throughput of NPU-only execution.
On VBench-2.0, \sysname matches FP32 in the total score over all 18 dimensions.
\end{abstract}



\received{20 February 2007}
\received[revised]{12 March 2009}
\received[accepted]{5 June 2009}

\maketitle

\section{Introduction}
\label{sec:intro}

\begin{figure}[t]
  \centering
  \includegraphics[width=\columnwidth]{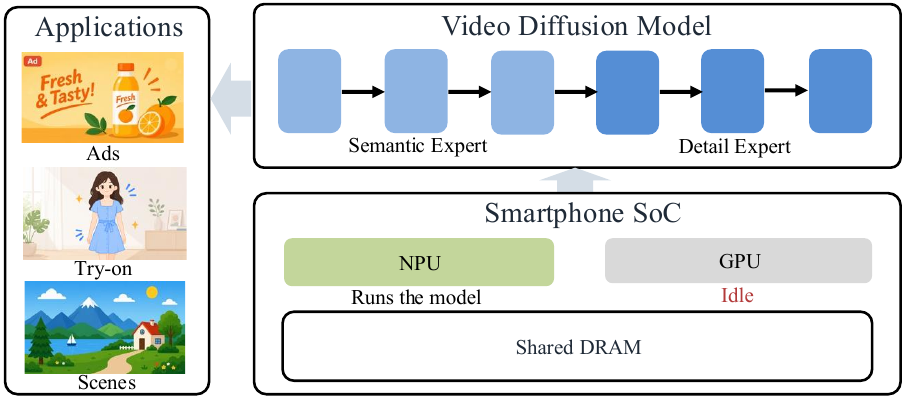}
  \caption{Applications that generate video from user-specific inputs run on a
  phone whose SoC carries both a programmable NPU and a GPU on shared DRAM.
  The workload is a multi-expert diffusion transformer that switches weight sets
  part-way along its denoising trajectory. Published on-device systems run
  that transformer on one engine.
  The other either sits idle or only decodes the finished latent afterwards.}
  \Description{Three panels. Left, four application icons: advertising,
  avatars, virtual try-on, and scene generation. Centre, a six-step denoising
  trajectory whose steps are coloured to show a switch of expert half-way.
  Right, a smartphone SoC with an active NPU, an idle GPU, and shared DRAM.}
  \label{fig:teaser}
\end{figure}

Generative video diffusion has progressed from closed research demonstrations to open model releases and product previews~\cite{openai2024sora,wan2025,hunyuanvideo2024,cogvideox2024,google2024veo}.
With these models, an application no longer selects a precomputed asset but synthesises a short clip for a specific user when it is needed.
Representative applications include personalised video advertising, audio-driven avatars, video virtual try-on, and interactive scene generation~\cite{kapoor2025video_ads,emo2024,karras2024fashion_vdm,yu2025wonderworld}.
Most of these applications need many clips at once.
An advertising feed renders one clip per product, an avatar one per audio segment, a try-on session one per shortlisted garment, and scene generation one per user action.
Such an application therefore waits for a whole set of clips, and it cares about how long the set takes rather than how long any single clip takes.
On a phone one clip already takes minutes, so the time for a set comes down to the rate at which the device completes clips.
That rate decides whether such an application is usable, and we call it \emph{stream throughput}.

The phone already has the hardware to raise this rate, but no published system uses all of it.
A modern smartphone integrates a programmable neural processing unit (NPU) and a graphics processing unit (GPU) that share the same physical DRAM~\cite{snapdragon8elite,hexagonv79,adreno830,nnstretch2023,heteroinfer2025}.
Yet on-device diffusion systems run the denoiser on one of them, as Figure~\ref{fig:teaser} shows.
Some works~\cite{snapgenv,mobilediffusion2024,mobilevd} shrink the model until it fits mobile constraints and run it on a single engine, and recent video systems~\cite{neodragon,mobilewan} place the whole diffusion transformer on the NPU.
One system~\cite{s2dit} uses the GPU only for the decoder, so the two engines run in sequence rather than together.
In every case the GPU stays outside the denoising computation.

The idle GPU matters because such sets of clips have strong reasons to be generated on the device rather than remotely.
When conditioning data contain personal information, remote processing raises privacy and compliance considerations~\cite{gdpr-2016,ccpa-2018,edge-privacy-survey-2021}.
Remote execution is exposed to network latency, and under load to queueing before denoising begins~\cite{wang2020edge,diffserve2025}.
At scale, centralised generative inference also incurs recurring cloud-GPU serving cost per clip~\cite{spotserve2024}.
Image diffusion and language models have already moved towards on-device execution~\cite{mobilediffusion2024,mlc-llm,heteroinfer2025}.
The video systems above follow the same path, but repeated transformer execution creates a different deployment problem for video.

Two properties of modern video diffusion make it hard to run on a phone.
First, one video executes $L$ transformer blocks over $N$ denoising steps, yielding $N{\cdot}L$ block executions against a shared latent.
Few-step distillation~\cite{salimans2022progressive,song2023consistency} reduces $N$ from tens to a few~\cite{causvid,t2vturbo}, but it does not remove this repeated structure.
Second, recent \emph{multi-expert} diffusion models specialise denoisers for different noise regimes~\cite{ediffi,ernievilg2,raphael,wan2025}.
For example, the Dual-expert Consistency Model (DCM)~\cite{dcm2025} freezes its semantic expert and adds timestep-dependent embeddings and a low-rank adaptation (LoRA) of the attention projections for detail refinement.
TimeStep Master also attaches timestep-specific LoRA experts to a shared base model~\cite{timestepmaster}.
Together, these properties create long accelerator occupancy and a large persistent weight footprint.
In a stream these costs are paid again for every clip. The time a clip occupies the accelerator sets how fast clips complete, and the weights must stay in memory for the whole stream.

A cross-engine system could use the two engines and the shared DRAM to address both costs.
With a stream of clips, the GPU can run part of the denoising computation while the NPU runs the rest, so both engines are busy at the same time.
Because they share physical DRAM, they can also read one stored copy of the weights instead of holding one each.
This does not make a single video faster, because its trajectory is serial and the GPU is slower per block, but it shortens the period between finished videos in the stream.
That period is the stream throughput the applications above wait on, so we target it rather than single-video latency.

Three deployment obstacles stand between this opportunity and a practical cross-engine system.
First, cross-engine execution requires a compatible stored weight representation without maintaining engine-specific converted copies.
Each accelerator runtime normally expects weights in its own layout, so adding an engine requires a separately converted representation.
Every converted copy adds to the memory the weights occupy, and multi-expert models add expert parameters on top.
Second, a partition must balance two asymmetric engines.
But the balance is not a simple ratio of block rates, because the decoder runs only on the NPU and adds a fixed per-video cost to its side.
Fine-grained GPU--NPU execution has been used for mobile LLMs, where HeteroInfer partitions each weight tensor so that both engines compute the same layer concurrently~\cite{heteroinfer2025}.
Applied to video diffusion, a boundary at every block execution would require $N{\cdot}L$ synchronisation points per video, each of which has to make the whole activation available to both runtimes before either engine proceeds.
Cutting only at step boundaries avoids the coordination but leaves too few places to balance two engines of very different speed.
Third, the platform has limits that no data sheet states.
How large one buffer may be, whether both runtimes can map the same buffer, and how much activation memory the NPU toolchain lets one graph hold are all capped, and the caps are found only by trying.
When both engines run at once they also compete for DRAM bandwidth and power, so a block takes a different time than running alone.
No existing system clears all three obstacles, so a multi-expert video model still has no exact cross-engine serving path on a phone.

We present \sysname, a system that serves a multi-expert video diffusion model across the NPU and the GPU of one phone, evaluated on a Snapdragon~8~Elite with a two-expert model.
It answers the three obstacles with three novel designs.
First, the base weights are stored once in INT8 using the layout the NPU reads natively.
The GPU uses the same bytes directly as operands of its integer dot-product instruction, so neither runtime rearranges the weights on load.
Both runtimes map the same physical pages, and the NPU dequantises the block it is executing to FP16 in a scratch buffer that the next block overwrites, so each engine keeps its own fastest arithmetic while the memory stores only one copy.
Second, the denoising trajectory is flattened into its $N{\cdot}L$ block executions and cut once at a block boundary, which may lie inside a step.
The GPU runs the block executions before the cut, and the NPU runs the rest and then the decoder, so in a stream the GPU starts video $i{+}1$ while the NPU finishes video $i$.
Every block still runs in order on the full activation, so the partition is algorithmically exact.
We place the cut from rates measured while both engines run together, with the decoder's per-video cost.
Third, the platform limits are measured first and designed around.
We measure how large a buffer can be, which buffers both runtimes can map, how much activation memory one graph may hold, and how much each engine speeds up or slows down when the other runs.
The size limits set how the weights are divided into buffers, the co-run rates set the cut, and the same rates predict the throughput.

We evaluate \sysname on a Galaxy~S25 running DCM-Wan2.1, measuring performance and scoring output quality.
On a stream of 49-frame videos, \sysname completes one every $677$~s and NPU-only every $917$~s, a stream-throughput gain of $1.35\times$. The gain rises to $1.50\times$ at 17 frames.
The gain holds under a service-level objective (SLO). \sysname sustains $5.3$ videos per hour against $3.9$ for NPU-only while every video meets the deadline.
The two-engine pipeline holds $1.47$~GiB of weights, half the $2.83$~GiB that data parallelism needs for its two working sets.
Quality is unchanged within $0.2$~pp of FP32 in VBench Quality score and within $1.3$~pp in VBench-2.0 score.

This paper makes three contributions:
\begin{itemize}\itemsep=2pt\parsep=0pt\topsep=3pt
\item \textbf{One weight copy for two engines.}
We store the base weights once in INT8 with the NPU's native layout, and the GPU consumes the same bytes directly.
Both runtimes map the same physical pages.

\item \textbf{One cut with no approximation.}
We flatten the trajectory into $N{\cdot}L$ block executions and cut it once at a block boundary.
Every block runs in order on the full activation.

\item \textbf{Measured platform limits as design inputs.}
We measure how large a buffer both runtimes can register, how much activation memory the NPU toolchain lets one graph hold, and how co-execution changes each engine's rate.
The design built from these numbers sustains $1.35$--$1.50\times$ the stream throughput of NPU-only execution on a Galaxy~S25.
\end{itemize}

\section{Background}
\label{sec:background}

\subsection{Video Diffusion and Multi-Expert Models}
\label{sec:bg:diffusion}

\begin{figure}[t]
  \centering
  \includegraphics[width=\columnwidth]{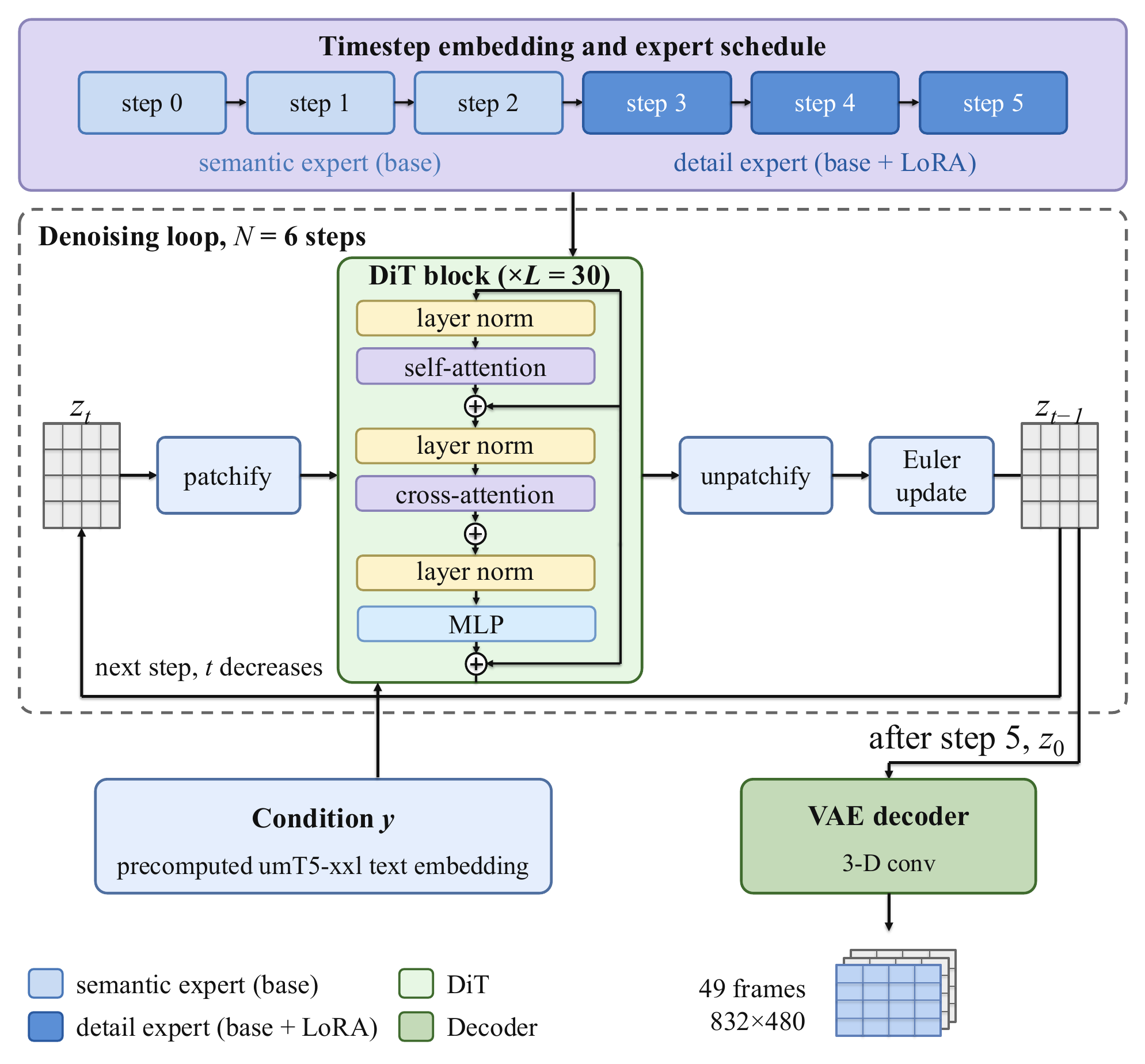}
  \caption{Inference in DCM-Wan2.1, the two-expert video diffusion model we
  deploy. Top: steps 0--2 use the semantic expert, which is the shared
  base, and steps 3--5 (the switch step $\kappa{=}3$) the detail expert as the same base with a rank-128 attention LoRA and its own timestep embedding. Middle: one step is
  one forward pass. The latent $z_t$ is patchified, passes $L{=}30$ DiT blocks whose cross-attention reads the precomputed umT5 text embedding $y$, and unpatchify and one Euler update give $z_{t-1}$. Bottom: after step 5 the variational-autoencoder (VAE) decoder, a stack of 3-D convolutions, turns $z_0$ into 49 frames at $832{\times}480$.}
  \Description{Flow diagram. A top band shows six step boxes, the first three
  in light blue labelled semantic expert and the last three in dark blue
  labelled detail expert, with an arrow from the band into the DiT block.
  Below, a dashed box holds the denoising loop: latent z t, patchify, a DiT
  block repeated 30 times with layer norm, self-attention, cross-attention and
  feed-forward network and three residual additions, unpatchify, an Euler update, and the next
  latent feeding back. A condition box with the precomputed umT5 text
  embedding feeds the DiT block from below. After step 5 the latent z 0 goes
  to a VAE decoder of 3-D convolutions, which outputs 49 frames at 832 by 480.
  A legend gives the two expert colours, the DiT colour and the decoder
  colour.}
  \label{fig:dcm}
\end{figure}

A video diffusion model denoises a Gaussian latent $z_t$ over $N$ timesteps~\cite{ho2020denoising,song2021ddim}.
Each step patchifies $z_t\in\mathbb{R}^{C_z\times F_z\times H_z\times W_z}$ with patch size $p$ into $\bar{z}_t\in\mathbb{R}^{M\times d_z}$ with $M=F_z(H_z/p)(W_z/p)$ tokens.
A 480p 49-frame clip gives $M{\approx}2{\times}10^{4}$.
The tokens pass $L{=}30$ Diffusion Transformer (DiT) blocks~\cite{peebles2023dit}, and block $\ell$ outputs $X_{t,\ell}\in\mathbb{R}^{M\times d}$.
An Euler update then yields $z_{t-1}$.
After the last step a 3-D convolutional VAE decoder produces the frames.

Multi-expert models specialise denoiser parameters by timestep.
Wan2.2-A14B~\cite{wan2025} switches by timestep between two $14$~B experts.
They need $28$~GB in INT8 and do not fit.
TimeStep Master~\cite{timestepmaster} attaches LoRA experts to one backbone.
DCM~\cite{dcm2025} shares parameters by construction (Figure~\ref{fig:dcm}).
The semantic expert is a $1{,}419$~M shared base.
The detail expert adds a rank-128 LoRA on the attention projections ($94.4$~M) and its own timestep-embedding block ($25.6$~M), and it takes over from step index $\kappa{=}3$ of six fixed timesteps.
We therefore serve DCM-Wan2.1.
Two properties matter.
A step is one forward pass with no second branch to overlap.
Six steps give too few partition points to balance the GPU and NPU, so §\ref{sec:design} cuts at block granularity.

\subsection{Snapdragon 8 Elite Silicon}
\label{sec:bg:silicon}

The Galaxy S25 uses the Snapdragon 8 Elite~\cite{snapdragon8elite}, whose two accelerators share DRAM.
Its Hexagon V79 NPU couples the HVX vector unit with the HMX matrix unit.
HMX multiplies $32{\times}32$ INT8 or FP16 tiles held in VTCM in a fixed tile order~\cite{hexagonv81hmx,snapdragon8elite}.
The NPU is reached over FastRPC~\cite{qualcomm-fastrpc} and runs offline-compiled QNN context binaries~\cite{qnn-sdk} with fixed shapes and an activation-footprint limit.
The Adreno 830 GPU~\cite{adreno830} runs OpenCL with INT8 \texttt{dp4a} and no matrix unit.
Usable memory is $10.85$~GiB.
The runtimes share memory only through DMA-BUF descriptors~\cite{linux-dmabuf} from \texttt{AHardwareBuffer}~\cite{android-ahb} or the Hexagon heap, which the CPU, GPU and DSP map at once.
QNN accepts memory only in this form.
Appendix~\ref{app:background} expands this section.

\section{Motivation}
\label{sec:motivation}

This section reports what we measured when we deployed a dual-expert video diffusion model on a Galaxy~S25.
The workload is DCM-Wan2.1~\cite{dcm2025} at its released configuration.
It generates 49 frames at $832{\times}480$ in $N{=}6$ denoising steps.
The diffusion latent has $C_z{=}16$ channels and $F_z{=}13$ temporal positions.
Its spatial dimensions are $H_z{=}60$ and $W_z{=}104$.
The DiT uses $p{=}2$ and $d{=}1536$, giving $d_z{=}64$ values per latent patch and $M{=}20{,}280$ tokens.
Each step runs $L{=}30$ DiT blocks.
The measurements expose three linked questions: why whole-video assignment fails, how both runtimes can access the same base-weight copy, and where the cut should fall.

\subsection{Whole videos are the wrong scheduling unit}
\label{ssec:mot:prize}

Video generation on a phone today keeps only one engine busy~\cite{neodragon,mobilewan,s2dit}.
As shown in Table~\ref{tab:bench_baselines}, our NPU-only pipeline produces one video every $916.7$~s with the decoder inside the pipeline, and the GPU is idle for all of it.
Almost half of the SoC's programmable accelerator does no work.

The obvious way to use both engines is to hand each of them whole videos, and that fails for two independent reasons.
First, at the solo rates in Table~\ref{tab:bench_baselines}, the $N{\cdot}L{=}180$ DiT block executions take $2{,}350$~s on the GPU and $603$~s on the NPU, a $3.90\times$ difference.
Second, the GPU cannot finish a video at all, because the Qualcomm AI Runtime (QAIRT) GPU backend defines no \texttt{Conv3d} and the decoder never runs there.
Using both engines therefore requires splitting each trajectory once between them.
The next feasibility question is whether both runtimes can access the model without duplicating its weights.

As shown in Table~\ref{tab:precision}, the two engines do not agree on a precision either.
The GPU runs a block fastest in INT8 and pays $3.37\times$ for FP16, while the NPU runs the same block fastest in FP16 and pays $2.96\times$ for INT8.
The GPU is fast in INT8 because its integer dot-product instruction does four INT8 multiplies at once, whereas floating point does one; FP16 does not help because the GPU converts every FP16 value to FP32 before computing on it anyway.
The NPU is fast in FP16 because its matrix unit computes on FP16 natively; INT8 is slower because the hardware cannot read signed INT8 results directly, so each GEMM needs an extra quantise pass before it and a dequantise pass after it, and those passes cost more than the GEMM itself.
A shared representation therefore cannot simply be whichever format each engine prefers.

\begin{table}[t]
\centering
\caption{Running the whole workload on one engine, alone and while the other
engine is busy. Co-execution moves the two engines in opposite directions. The NPU solo figure is the mean of ten videos generated back to
back with decoding inside the pipeline.}
\label{tab:bench_baselines}
\begin{tabular}{lrrr}
\toprule
 & per DiT block & VAE & per video \\
\midrule
NPU, solo   & $3{,}352$~ms  & $308$~s & $916.7$~s \\
NPU, co-run & $3{,}101$~ms  & $275$~s   & $825$~s   \\
GPU, solo   & $12{,}633$~ms & ---       & ---       \\
GPU, co-run & $12{,}800$~ms & ---       & ---       \\
\bottomrule
\end{tabular}
\end{table}

\begin{table}[t]
\centering
\caption{One DiT block on each engine at three precisions, same block and same
input at $M{=}20{,}280$. Each multiplier is against that engine's own fastest
precision, shown in bold. The two engines prefer opposite formats.}
\label{tab:precision}
\begin{tabular}{lrrr}
\toprule
engine & INT8 & FP16 & FP32 \\
\midrule
GPU & $\mathbf{12.6}$~\textbf{s}   & $42.3$~s ($3.37\times$) & $37.8$~s ($3.01\times$) \\
NPU & $9.95$~s ($2.96\times$)      & $\mathbf{3.36}$~\textbf{s} & $3.95$~s ($1.17\times$) \\
\bottomrule
\end{tabular}
\end{table}

\subsection{Shared weights require per-block regions}
\label{ssec:mot:memory}

Running two engines naively duplicates the weights.
Each runtime stages tensors in its own layout, so one engine-native FP16 copy occupies $2.82$~GB and two copies occupy $5.64$~GB.
That is about half of the $11.65$~GB the kernel reports usable, and everything else on the phone has to fit in what is left.

The obvious repair is to keep one base-weight copy that both runtimes read.
The platform will not allow us to allocate it naively.
Adreno advertises a maximum OpenCL buffer of $1024$~MiB, but Android's gralloc refuses a BLOB allocation at exactly $2^{29}$~B, so OpenCL never gets the chance to refuse.
A $511$~MiB region registers into a QNN Hexagon context and an OpenCL context at the same time, while $1024$~MiB takes the device down.
The design that survives divides the shared base weights into one region per DiT block, with every region below the verified limit.
This resolves the storage constraint; the remaining question is how to divide the trajectory between the two engines.

\subsection{The cut must be block-level and measurement-driven}
\label{ssec:mot:ratio}

A single handoff divides each trajectory into a GPU prefix and an NPU suffix.
The NPU suffix also runs the VAE decoder, so the cut must balance the complete pipeline rather than the DiT alone.

With $N{=}6$, a step-level cut moves work only in units of $L{=}30$ block executions.
At the co-run rates in Table~\ref{tab:bench_baselines}, the neighbouring step-level cuts occur after blocks $30$ and $60$.
The first leaves the GPU idle for $356$~s, while the second leaves the NPU idle for $121$~s.
A block-level cut can be placed between them and still requires only one handoff per video.

Selecting that block boundary requires the co-run GPU block time $T_G$, NPU block time $T_N$, and NPU decoder time $T_V$.
These rates depend on the workload, concurrent execution, and thermal state.
The cut must therefore be derived from measurements under the target deployment conditions.

Together, these measurements motivate a single-cut pipeline.
Its design must satisfy three requirements.
First, the shared base weights must be allocated at per-block granularity because a monolithic cross-runtime allocation is unavailable.
Second, the trajectory must use one block-level cut because step-level cuts cannot balance the two pipeline stages.
Third, the cut must be derived from co-run $T_G$, $T_N$, and $T_V$ measured under a stated cooling condition.
Section~\ref{sec:design} develops the resulting design, and Section~\ref{sec:eval} reports the measurements under these platform conditions.

\section{System Design}
\label{sec:design}

\begin{figure*}[t]
\centering
\includegraphics[width=\textwidth]{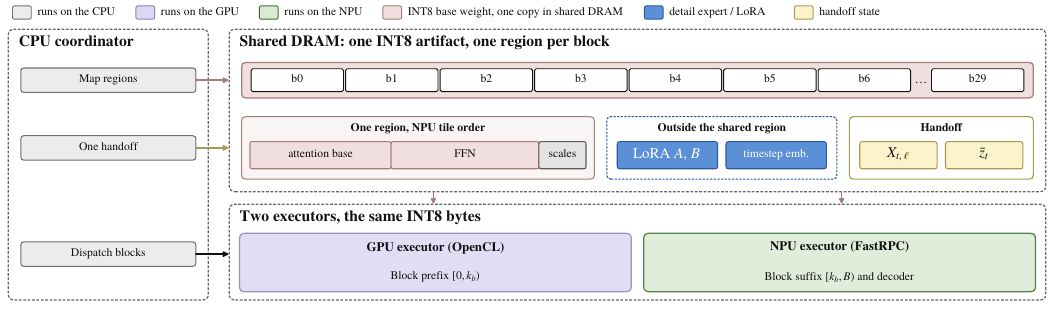}
\caption{\sysname in one view.
The CPU coordinator (left) maps each per-block weight region into both runtimes at start-up, dispatches the block prefix $[0,k_b)$ to the GPU and the suffix $[k_b,B)$ to the NPU, where $B{=}N{\cdot}L$ counts the block executions of one video, and performs the single hand-off, whose hidden state $X_{t,\ell}$ and latent $\bar{z}_t$ live in the same shared DRAM (lavender).
Both executors (bottom right) read the same INT8 bytes: the GPU uses them as integer dot-product operands and rescales in the epilogue, while the NPU dequantises the feed-forward weights once per block into an FP16 scratch and applies the adapter factors inside its block kernel.
The base is stored once in shared DRAM (top right), one region per block in the NPU's tile order; the detail expert's factors and timestep embedding live outside it.}
\Description{A wide diagram with a colour legend on top. Left, a CPU coordinator column of three grey boxes: map regions and one hand-off, each with an arrow into the shared DRAM, and dispatch blocks with an arrow into the executors. Top right, a pink shared-DRAM bar divided into thirty regions; below it a zoom into one region with attention-base, feed-forward and scale segments, a dashed box holding the LoRA factors and timestep embedding outside the shared region, and a yellow hand-off box holding the hidden state and the latent. Bottom right, a GPU executor box and an NPU executor box side by side, each with a one-line summary of how it consumes the INT8 bytes. Two red arrows lead from the DRAM down to both executors.}
\label{fig:overview}
\end{figure*}

\subsection{Overview}
\label{ssec:design:overview}

\sysname shares one physical copy of the base weights in shared memory between the GPU and NPU, then executes each denoising trajectory across both accelerators.
Each video requires one cross-engine handoff.
Figure~\ref{fig:overview} shows the system: a CPU coordinator, the two executors that consume one INT8 artifact, and the shared DRAM that holds it; Figure~\ref{fig:partition} shows how the trajectory is cut between the executors.

\sysname treats the $B=N{\cdot}L$ DiT block executions of one video as a single flattened sequence.
The GPU executes the prefix $[0, k_b)$ DiT blocks, while the NPU executes the suffix $[k_b, B)$ and decodes the final latent with the VAE.
Both engines read the same stored base-weight artifact.
It contains INT8-quantised weights in the NPU's native HMX tile layout and one FP32 scale per output column.
The GPU consumes these weights directly as \texttt{dp4a} operands.
The NPU uses FP16 operands for its GEMMs and caches the dequantised feed-forward weights for the current block (§\ref{ssec:design:sharedw}).
The GPU writes the hidden state $X_{t,\ell}\in\mathbb{R}^{M\times d}$ and the patchified running latent $\bar{z}_t\in\mathbb{R}^{M\times d_z}$ at the cut.
The NPU then resumes inside the same denoising step (§\ref{ssec:design:partition}).
The selected cut also falls before the first expert switch, so the GPU runs only the semantic expert and the NPU handles the switch locally (§\ref{ssec:design:partition}).
The GPU GEMMs use INT8 operands, while the NPU GEMMs use FP16 operands (§\ref{ssec:design:sharedw}).
The GPU keeps the state between blocks in FP16 and uses a separate scale for each GEMM input.
These choices limit error accumulation across the quantised block sequence (§\ref{ssec:design:precision}).
At each denoising step, the host CPU projects the current latent into DiT tokens, maps the final hidden state back to a latent-space prediction, and applies the Euler update to obtain the latent for the next step.
The mechanisms assume only a stack of DiT blocks whose base weights quantise to INT8 and a denoising trajectory that repeats those blocks on one latent, so $N$ and $L$ are free in the flattened sequence.
Neither assumption involves DCM or its LoRA experts.

Three rules govern the design, and each is forced by a measurement rather than chosen for elegance.

\begin{enumerate}\itemsep=1pt\parsep=0pt\topsep=2pt
\item \textbf{Share the weight representation, not the compute format.}
Both engines read the same INT8 base-weight artifact.
The GPU uses the weights directly for INT8 dot products, whereas the NPU dequantises them for FP16 computation (§\ref{ssec:design:sharedw}).
\item \textbf{Share weights one DiT block at a time.}
The platform cannot create one cross-runtime buffer large enough for the full base model.
\sysname therefore allocates the shared base weights at per-DiT-block granularity (§\ref{ssec:design:sharedw}).
\item \textbf{Cut once, at a block boundary.}
A step-level cut offers only $N{+}1$ positions.
Handoffs after every DiT block add repeated communication, so \sysname instead places one block-level cut in the flattened sequence (§\ref{ssec:design:partition}).
\end{enumerate}

\subsection{Cross-Engine Single-Copy Base-Weight Storage}
\label{ssec:design:sharedw}

Section~\ref{sec:motivation} established two constraints on weight storage; Figure~\ref{fig:overview} shows the artifact that satisfies them, how each executor consumes it, and what stays outside it.
Each shared region must fit within the platform limit, and the same stored representation must support INT8 GPU GEMMs and FP16 NPU GEMMs.
\sysname therefore uses one canonical INT8 base-weight artifact and divides it into one region per DiT block.

For a weight matrix $W\in\mathbb{R}^{d_{\mathrm{in}}\times d_{\mathrm{out}}}$, \sysname assigns one INT8 scale to each output column:
\begin{equation}
\begin{aligned}
s_j &= \frac{\max_i |W_{ij}|}{127},\\
Q_{ij} &= \operatorname{clip}_{[-127,127]}\!\left(\operatorname{round}\!\left(\frac{W_{ij}}{s_j}\right)\right).
\end{aligned}
\end{equation}
It then stores $Q$ in the $32{\times}32$ tiled layout required by the Hexagon matrix unit.
Tiles are ordered first by output-column tile and then by input-row tile.
The byte address of $Q_{ij}$ is
\begin{equation}
\begin{aligned}
\mathrm{off}(i,j) ={}&
\left(\left\lfloor\frac{j}{32}\right\rfloor\frac{d_{\mathrm{in}}}{32}
+ \left\lfloor\frac{i}{32}\right\rfloor\right)1024 \\
&+ \left\lfloor\frac{i \bmod 32}{4}\right\rfloor128
+ (j \bmod 32)4 + (i \bmod 4).
\end{aligned}
\label{eq:hmx}
\end{equation}
Each DiT block is stored in a separate region that satisfies the allocation limit measured in Section~\ref{ssec:mot:memory}.

The same tiled representation is consumed differently by the two engines.
The last term in Equation~\ref{eq:hmx} places four consecutive input values for a fixed output column in four contiguous bytes, which is exactly the \texttt{char4} operand Adreno's integer dot-product instruction takes.
The GPU reads these bytes directly and applies the activation and weight scales in the GEMM epilogue as $y_j=\mathrm{acc}_j(s_xs_j)+b_j$, where $\mathrm{acc}_j$ is the integer accumulator, $s_x$ the calibrated activation scale of §\ref{ssec:design:precision}, and $b_j$ the output bias.
For the feed-forward layer, the NPU dequantises both weight matrices to FP16 once at the start of each block.
All feed-forward chunks in that block reuse the block-local FP16 scratch buffer.
The remaining weights follow their kernel-specific staging paths.

The NPU's HMX tile layout is the canonical storage format because the NPU requires this layout in hardware.
The GPU can support it by changing its OpenCL index calculation.
A GPU-native row-major layout would instead make the NPU reorder the same weights repeatedly while staging a block.

At runtime, the GPU and NPU map each per-block region into their own address spaces while referring to the same physical pages in shared memory.
The two runtimes therefore share one physical copy of the base weights.
Their runtime handles and virtual addresses remain separate, but no second base-weight allocation is created.
This removes cross-engine duplication of the base weights.
Expert-specific parameters and the NPU's temporary FP16 scratch remain separate.

The factorised expert representation stores the shared attention base $W$, the rank-128 LoRA factors $A$ and $B$, and the expert-specific timestep embeddings separately.
The detail expert evaluates
\begin{equation}
y=xW+(xA^{\top})(2B)^{\top},
\end{equation}
while both experts reuse the same feed-forward weights.
Keeping the factors separate avoids storing a second full-rank attention copy and avoids requantising a small adapter after it is merged into the base.
Both device runners consume this representation at run time.
The NPU runs the two factor GEMMs inside its block kernel and adds the result to the base projection in FP16.
The GPU computes the same low-rank delta and fuses it into the GEMM epilogue before requantisation.
Merging the adapter into an INT8 base is not merely unnecessary: the adapter's magnitude is about one third of the base's requantisation noise, so a merged INT8 copy cannot represent the two experts distinguishably at weight level, whereas the factorised path applies the delta at full precision.
The factorised expert representation is the only part of \sysname that is specific to DCM.
A single-expert model omits it, and a model whose experts each carry a full base stores one shared artifact per expert under the same design.

The NPU reuses its feed-forward scratch block by block.
The temporary FP16 copy is therefore limited to the block currently executing rather than the full model.
Section~\ref{sec:eval} reports the memory, latency, and fidelity effects of these choices.

\subsection{Partitioning the Flattened Trajectory}
\label{ssec:design:partition}

\begin{figure}[!htb]
\centering
\includegraphics[width=\columnwidth]{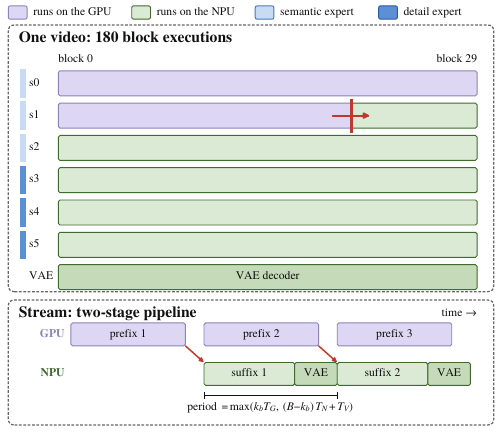}
\caption{Partitioning the flattened trajectory.
Top: one video is $B{=}180$ block executions followed by the VAE decode, cut once at $k_b{=}51$ inside step~1 (red). The GPU runs the prefix, the NPU the suffix and the decoder; only the hidden state and the running latent cross, once per video, and because $k_b{\leq}\kappa L{=}90$ the expert switch stays on the NPU.
Bottom: over a stream, the GPU prefix of the next video overlaps the NPU suffix and decode of the current one, so the period is the longer of the two stages (Equation~\ref{eq:kb}); the decoder's fixed cost on the NPU moves the balance point later than the block rates alone would place it.}
\Description{Two stacked panels. Top, six rows of thirty blocks, coloured lavender where the GPU executes and green where the NPU executes, with a red cut and arrow at block 51 in step 1 and a light-blue or dark-blue bracket per step for the expert. Bottom, a two-row timeline: the GPU row holds the prefixes of videos 1 to 3, the NPU row the suffix and decode of videos 1 and 2 shifted by one period, with red handoff arrows and a bracket marking one period.}
\label{fig:partition}
\end{figure}

To obtain finer placement than a step-level cut without introducing repeated handoffs, \sysname flattens the trajectory into $B=N{\cdot}L$ block executions and places one block-level cut at index $k_b$ (Figure~\ref{fig:partition}).
The GPU executes the prefix $[0,k_b)$.
The NPU executes the suffix $[k_b,B)$ and decodes the final latent, so no additional handoff is needed.

For a stream of videos, the GPU prefix and NPU suffix form a two-stage pipeline.
Let $T_G$ and $T_N$ denote the GPU and NPU time per DiT block, and let $T_V$ denote the NPU time for VAE decoding.
The steady-state period is
\begin{equation}
\mathrm{period}(k_b) = \max\big(k_b\,T_G,\; (B{-}k_b)\,T_N + T_V\big).
\label{eq:kb}
\end{equation}
The two terms cross at
\begin{equation}
k = \frac{B\,T_N + T_V}{T_N + T_G}.
\end{equation}
The integer optimum $k_b^{*}$ is the neighbouring integer that gives the smaller period.
The decoder cost appears only in the NPU stage, so omitting $T_V$ would balance the DiT blocks rather than the complete video pipeline.

The three latency values are not fixed properties of the hardware.
\sysname measures $T_G$, $T_N$, and $T_V$ while both engines run under the target workload and thermal condition.
It then evaluates the neighbouring integer cuts and selects the one with the lower predicted period.
The current runner fixes $k_b$ at startup and does not adjust it online.

Only the state needed to resume the current denoising step crosses at the cut.
The GPU hands off the hidden state $X_{t,\ell}$ and the patchified running latent $\bar{z}_t$ used by the Euler update.
Because $k_b$ need not be a multiple of $L$, the NPU may resume in the middle of a denoising step.
Its block kernel accepts a starting block and a block count, so it skips the patch embedding and loads the transferred hidden state.

The offline artifact stores the metadata needed to resume execution on the NPU, including the AdaLN parameters, expert choice, and Euler step size for each denoising step.
The GPU therefore hands off only the two state tensors.
The cut index $k_b$ is a runtime parameter selected at startup; the artifact stores only the per-step metadata needed to resume at any block boundary.

The cut also determines whether expert routing crosses the engine boundary.
The expert switch remains on the NPU when $k_b \leq \kappa L$, where $\kappa L$ is the first block that uses the detail expert.
The GPU therefore runs only the semantic expert, while the NPU selects the detail expert locally without moving expert parameters.
During artifact generation, \sysname records the expert used by each timestep so that routing remains correct when the timestep schedule changes.

\subsection{Activation Numerics and Calibration}
\label{ssec:design:precision}

The GPU keeps the hidden state between DiT blocks in FP16, the last row of the GPU executor in Figure~\ref{fig:overview}.
Before each INT8 GEMM, it quantises the GEMM input using a scale calibrated for that block and input.
This allows the GPU to use INT8 dot products without keeping intermediate states in INT8.
The NPU keeps the hidden state between blocks in FP32 and narrows it to FP16 only where it enters a GEMM, because the state accumulates 30 gated additions per step and grows to about $25$ in magnitude by the last block, and it never serves as a GEMM operand, so the wider format costs nothing.

Before deployment, we calibrate each GEMM input using the actual timestep schedule and AdaLN modulation. For each block and GEMM input, we divide the largest observed absolute value by 127 to obtain a fixed INT8 scale. The same scale is then used throughout inference.
We follow prior timestep-aware calibration methods~\cite{ptq4dm,qdiffusion,tdq,viditq}.

These choices reduce quantisation error but do not make INT8 execution identical to full-precision execution.
Section~\ref{sec:eval} evaluates its effect on generated-video quality.

\section{Implementation}
\label{sec:impl}

This section follows the execution path from runtime setup through the DiT
kernels and VAE decoder, then describes artifact generation and validation.

\sysname{} is about $25$\,k lines of new code. The NPU kernels are C with HVX and HMX intrinsics, added to the open-source \texttt{htp-ops-lib} operator package~\cite{hao2026htp}. The GPU executor and the coordinator are C++ and OpenCL C. Artifact export, quantisation and validation are Python, and device-side orchestration is shell.

\subsection{Runtime and Work Placement}
\label{ssec:impl:runtime}

\sysname consists of a device-side coordinator, an OpenCL GPU DiT executor, and a FastRPC NPU DiT executor.
The coordinator assigns the block executions before the cut to the GPU executor and the rest to the NPU executor, launches them for each video, and sequences the one handoff between them.
At the handoff the GPU passes two tensors, the hidden state and the running latent, and the NPU resumes at block $k_b$ even when the cut falls inside a denoising step.
Each executor runs the per-step patch embedding, output head and Euler update on the CPU in its own process, in place on the shared buffers the accelerator has just written, without copying them.
Both DiT executors bypass the QAIRT graph runtime.
The NPU VAE decoder uses the QNN Hexagon Tensor Processor (HTP) backend separately after the last block of each video, so in a stream the GPU prefix of video $i{+}1$ overlaps the NPU suffix and the decode of video $i$.

The NPU DiT executor calls a custom Hexagon operator package directly.
The QNN HTP backend prevents external operators from using HMX.
It also cannot serialise the full attention graph at the deployment shape without slicing it.
The GPU DiT executor uses hand-written OpenCL because the QNN GPU backend does not support the required fully connected operator.
Both executors expose the same block-range interface and read the artifact described in §\ref{ssec:design:sharedw}.

At startup, each block's base weights are placed in a shared DMA-BUF.
The NPU maps the buffer through FastRPC.
The GPU imports the same buffer through Qualcomm's OpenCL host-pointer extension.
When the two executors run as separate processes, the NPU-side owner passes the per-region file descriptors to the GPU process over a Unix-domain socket.
The runtimes use different handles and virtual addresses for the same physical pages in shared memory.
The rank-128 adapters, detail-expert timestep embeddings, and the NPU's block-local FP16 scratch remain separate from this shared base.
Appendix~\ref{app:impl} describes the NPU and GPU DiT kernels and the VAE decoder.

\begin{figure*}[t]
\centering
\includegraphics[width=\textwidth]{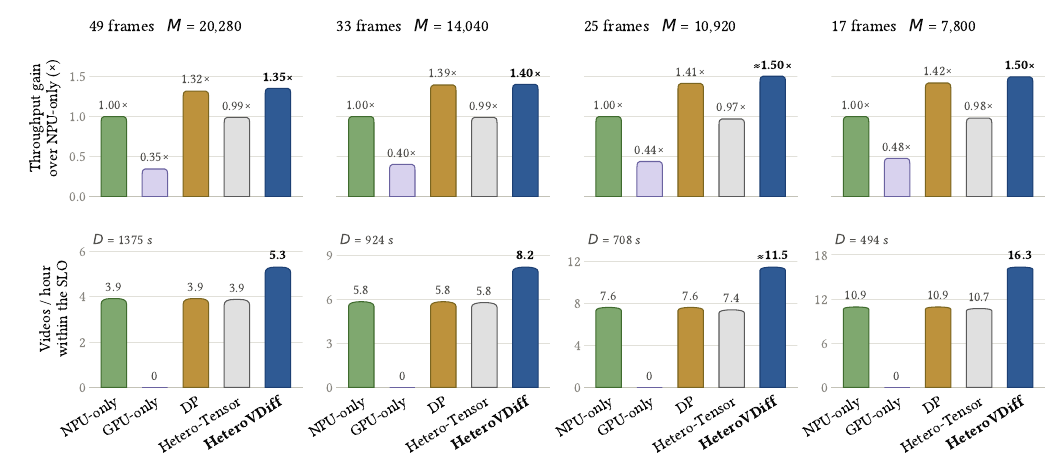}
\caption{Every configuration at four clip lengths.
Top: stream throughput as the gain over NPU-only (the vertical line at $1{\times}$ is the NPU-only baseline).
Bottom: videos per hour with every video inside the SLO,
a per-video deadline $D$ of $1.5{\times}$ the NPU-only latency.
DP drops to the NPU-only bar because its GPU-arm videos exceed the deadline $D$ and only its NPU arm counts.
Same device, active cooling, shared-base plus factorised-expert weight path, ten-video streams.}
\Description{Two rows of four vertical bar panels, one column per clip length
from 49 to 17 frames, five configurations per panel. Top row: the HeteroVDiff bar,
in blue, is the tallest in every panel at 1.35, 1.40, 1.50 and 1.50 times
NPU-only; DP in dark gray is just behind; Hetero-Tensor sits slightly below the
1.0 line; GPU-only is far below at 0.35 to 0.48. Bottom row: HeteroVDiff reaches
5.3, 8.2, 11.5 and 16.3 videos per hour within the deadline, NPU-only, DP and
Hetero-Tensor tie at 3.9, 5.8, 7.6 and 10.9, GPU-only is zero.}
\label{fig:throughput}
\end{figure*}





\section{Evaluation}
\label{sec:eval}

\subsection{Experimental Setup}
\label{ssec:eval:setup}

\paragraph{Platform and workloads.}
All device measurements use one Samsung Galaxy~S25 (Snapdragon~8~Elite, SM8750) with the Hexagon~V79 NPU and Adreno~830 GPU, and the usable memory is $10.85$~GiB as measured.
The model is DCM-Wan2.1-1.3B with $N{=}6$ denoising steps, $L{=}30$ blocks per step, and $832{\times}480$ output. The semantic expert on steps $0$--$2$ and the detail expert on steps $3$--$5$ ($\kappa{=}3$).
We vary the clip length over 49, 33, 25 and 17 frames, which sets the token count to $M{=}20{,}280$, $14{,}040$, $10{,}920$ and $7{,}800$ while the number of block executions per video stays $B{=}N{\cdot}L{=}180$.
Prompt embeddings are computed once offline, so the text encoder is not resident during generation.

\paragraph{Thermal and power protocol.}
Every timed run is \emph{cold}.
The phone sits on an external cooler with the screen off and stays plugged in.
Both controls hold the device in one operating state, so that the differences we report come from the schedule rather than from heat or a draining battery.

\paragraph{Metrics and measurement.}
The primary metric is \emph{stream throughput}, reported as seconds per video over a stream of 10 content-diverse prompts generated back to back with decoding inside the pipeline.
For the two-engine pipeline it is the steady-state period after the initial fill.
We also report single-video latency from launch to finished decode, SLO attainment and resident memory.
The SLO is a per-video deadline of $1.5\times$ the NPU-only single-video latency at the same workload.
Output quality is scored on the seven VBench quality dimensions, and on 18 VBench-2.0 dimensions.

\paragraph{Configurations.}
All self-built configurations consume the same INT8 base-weight artifact and apply the factorised expert representation of §\ref{ssec:design:sharedw} at run time, so they differ only in how execution is divided between the two engines.
\emph{NPU-only} runs the whole trajectory and the decoder on the NPU. It is the strongest single engine and serves as the ($1{\times}$) reference throughout.
\emph{GPU-only} runs the DiT on the GPU and decodes on the NPU, because the vendor GPU backend has no 3D convolution and therefore cannot run the VAE decoder.
\emph{DP} is naive data parallelism in which both engines generate whole videos from one prompt queue with separate working sets. The NPU worker decodes the GPU worker's latents as they arrive.
\emph{Hetero-Tensor} is the per-GEMM column split of HeteroInfer~\cite{heteroinfer2025} in which both engines compute a slice of every GEMM and exchange results after each one.
HeteroInfer has no public code, so we implemented the split ourselves following the paper.
\sysname{} places one block-level cut at the $k_b$ that Equation~\ref{eq:kb} selects from co-run rates for each workload, namely $51$, $54$, $59$ and $61$.

\subsection{End-to-End Throughput and SLO}
\label{ssec:eval:throughput}

Figure~\ref{fig:throughput} plots the gain over NPU-only for every configuration at all four workloads.
\sysname raises stream throughput over NPU-only at every clip length.
At 49 frames it completes a video every $677$~s against $917$~s, which is $5.3$ against $3.9$ videos per hour, a gain of $1.35{\times}$.
The gain rises to $1.50{\times}$ at 17 frames.
The deployed cuts are the ones the rate model selects for each workload, $k_b{=}51$, $54$, $59$ and $61$, all before the first expert switch at block $\kappa L{=}90$.

Naive data parallelism comes closest on throughput and loses on the SLO axis.
It reaches $1.32$--$1.42{\times}$, $0.6$--$6.4\%$ behind \sysname{} at the four workloads.
Under the SLO every video DP adds over NPU-only misses the deadline.
Each DP video is generated entirely by one engine, so the added videos all come from the GPU arm.
A GPU-arm video takes $2{,}620$~s at 49 frames against the deadline of $1{,}375$~s.
The added videos are therefore unusable, and DP's SLO-compliant rate is its NPU arm alone.
That is the NPU-only rate, $3.9$ videos per hour against \sysname's $5.3$.
\sysname meets the deadline because every video crosses both engines and none takes longer than $1{,}325$~s.
Hetero-Tensor finishes a video in $860.5$~s at 49 frames, $0.94{\times}$ NPU-only, because the per-GEMM split saves about $50$~ms of GEMM time per block and pays $155$~ms of waiting and cache maintenance, so the $4{,}680$ hand-offs per video cost more time than the GPU's slices save.
GPU-only completes at $0.35$--$0.48{\times}$.

The gain grows as clips shorten, and the cut has to be placed from co-run rates (\S\ref{ssec:eval:ablation}).
As $M$ falls, the GPU-to-NPU block-time ratio drops from $3.9$ to $2.7$, so the balanced cut moves later and the GPU carries more of each video, and the NPU's co-run speed-up grows from $10\%$ to $16$--$19\%$.



\subsection{Memory Analysis}
\label{ssec:eval:memory}

\begin{table}[t]
\centering
\caption{Weight memory of each configuration at steady state. Base is the
shared attention base, the feed-forward pair and the scales, and adapter is
the rank-128 factors. Transient is the NPU's FP16 copy of the current block's
feed-forward pair and the GPU's staged block. Total is their sum in GiB.}
\label{tab:memory}
\footnotesize
\setlength{\tabcolsep}{3.5pt}
\resizebox{\columnwidth}{!}{%
\begin{tabular}{@{}lcrrr@{}}
\toprule
Configuration & Copies & Base + adapter & Transient & Total \\
 & & (MiB) & (MiB) & (GiB) \\
\midrule
FP16, one copy per engine & 2 & 5,379 & -- & 5.25 \\
Merged-expert INT8, one per engine & 2 & 3,746 & 52.5 & 3.71 \\
\midrule
NPU-only & 1 & 1,423 & 52.5 & 1.44 \\
GPU-only & 1 & 1,512 & -- & 1.48 \\
DP & 2 & 2,846 & 52.5 & 2.83 \\
Hetero-Tensor (column slices) & 1 & 1,423 & 52.5 & 1.44 \\
\sysname{} (shared pages) & 1 & 1,423 & 78.8 & 1.47 \\
\bottomrule
\end{tabular}}
\end{table}

Sharing one stored copy of the base weights between the two engines is what saves the memory.
\sysname{} holds $1.47$~GiB of weights.
Data parallelism holds two artifacts, $2.83$~GiB.
One copy per engine needs $3.71$~GiB in merged INT8 and $5.25$~GiB in FP16.
Table~\ref{tab:memory} counts tensor sizes because resident-set counters move by tens of megabytes with page cache and compressed swap.
Both engines map the same artifact pages.
The GPU reads the attention base in place and stages only the current block's feed-forward pair.
The deployed cut keeps every detail-expert block on the NPU.
Adding the GPU to NPU-only therefore costs only $26.3$~MiB.
The kernel charges the shared attention base half to each process.
A GPU that loaded the same weights from the file would hold a private copy of about $540$~MB.

The factorised expert representation stores each DiT block once in $47.3$~MiB, so the dual-expert artifact is $1.39$~GiB.
Of that $1{,}423$~MiB the attention base is $541.4$~MiB, the feed-forward pair $787.5$~MiB, the scales $2.6$~MiB and the rank-128 factors $91.5$~MiB.
The NPU's FP16 copy of one block's feed-forward pair adds $52.5$~MiB.
Merging the detail expert into a second attention copy would take $1.83$~GiB and one FP16 copy per engine $2.82$~GB (§\ref{ssec:mot:memory}).
The expert switch is also free.
Applying the factors at run time moves no base weights, whereas merged weights would reload the whole attention base in every video.

Working buffers add about $0.4$~GiB.
Two hidden-state carriers each hold $59.4$~MiB in FP16 for a 49-frame clip.
The GPU's de-fused attention keeps one $256$-row block of scores and probabilities, $178$~MiB at $M{=}20{,}280$, growing with $M$ rather than $M^2$.
The hand-off writes the hidden state and latent once per video, $130$~MB.
\sysname{} needs about $1.9$~GiB at steady state, leaving the rest of the $10.85$~GiB the kernel reports usable to the decoder and the application.



\subsection{Output Quality}
\label{ssec:eval:quality}
\begin{figure*}[t]
\centering
\includegraphics[width=\textwidth]{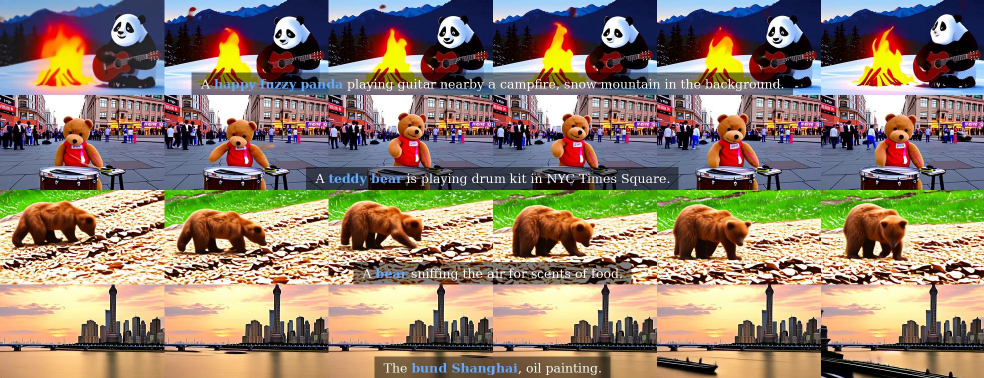}
\caption{Qualitative samples. Six uniformly spaced frames from each of four
$832{\times}480$, 49-frame videos generated under the deployed \sysname{} numeric
path ($k_b{=}51$) for VBench prompts, printed on each row.}
\Description{Four rows of six video frames each. Row one: a fuzzy panda playing
a guitar beside a campfire with snow mountains behind. Row two: a teddy bear
playing a drum kit in Times Square. Row three: a bear sniffing the air in a
field. Row four: the Bund in Shanghai rendered as an oil painting. Each row's
prompt is printed across its bottom edge.}
\label{fig:samples}
\end{figure*}

\begin{table}[t]
\centering
\caption{VBench~\cite{vbench} quality dimensions and Quality score (\%) of the four numeric paths defined in the text, full prompt suite with five samples per prompt.}
\label{tab:vbench}
\footnotesize
\setlength{\tabcolsep}{2.4pt}
\newcommand{\vbrot}[1]{\makebox[0pt][c]{\rotatebox[origin=l]{60}{#1}}}
\resizebox{\columnwidth}{!}{%
\begin{tabular}{@{}l*{7}{c}c@{}}
\toprule
Numeric path & \vbrot{Subject consistency} & \vbrot{Background consistency} & \vbrot{Temporal flickering} & \vbrot{Motion smoothness} & \vbrot{Dynamic degree} & \vbrot{Aesthetic quality} & \vbrot{Imaging quality} & \vbrot{\textbf{Quality score}} \\
\midrule
FP32 & 95.10 & 96.34 & 99.49 & 97.35 & 73.61 & 58.30 & 65.92 & 83.19 \\
NPU path & 95.02 & 96.29 & 99.48 & 97.31 & 75.28 & 58.09 & 66.00 & 83.25 \\
GPU path & 94.52 & 96.07 & 99.34 & 97.22 & 74.17 & 56.62 & 66.28 & 82.74 \\
\sysname, $k_b{=}51$ & 94.67 & 96.05 & 99.34 & 97.27 & 75.83 & 57.38 & 66.20 & 83.02 \\
\bottomrule
\end{tabular}}
\end{table}

We score four numeric paths of the same model, prompts and six-step schedule, which differ only in the arithmetic of the DiT blocks.
FP32 is the unquantised reference.
The NPU path consumes the shared INT8 base weights dequantised to FP16 and computes in FP16, the GPU path consumes the same INT8 weights with INT8 activations, and \sysname{} applies the GPU path to block executions $[0,51)$ and the NPU path to the remaining 129, matching the deployed cut.
Figure~\ref{fig:samples} shows frames from the deployed path, and Table~\ref{tab:vbench} scores all four paths on the full VBench prompt suite, while Table~\ref{tab:vbench2} in Appendix~\ref{app:vbench2} does the same on VBench-2.0.

The shared INT8 artifact costs no measurable quality.
On VBench the NPU path is within $0.21$\,pp of FP32 on six of the seven dimensions and $+0.06$\,pp in Quality score, the GPU path stays within $1.7$\,pp on every dimension and loses $0.45$\,pp, and \sysname{} sits between them at $-0.17$\,pp.

VBench-2.0, which judges faithfulness with vision--language-model (VLM) and detector-based scorers rather than perceptual quality, does not separate the four paths.
The Total scores lie within $1.3$\,pp of each other and none of the quantised arms falls below FP32.
Per-dimension differences run in both directions on every arm, and the largest, $+11.5$\,pp on multi-view consistency, is on a dimension the benchmark scores from only 8 to 10 videos per arm, so one video moves it by 10\,pp or more.
We therefore read VBench-2.0 as evidence of no separation at this prompt count, not of an improvement from quantisation.

\subsection{Ablations}
\label{ssec:eval:ablation}

\subsubsection{Where to cut}
As shown in Table~\ref{tab:cut}, a cut at a block boundary beats a cut at a step boundary because no step boundary reaches the balance point.
With $N{=}6$ the only step boundaries are multiples of $L{=}30$.
The two that bracket the deployed cut each leave one engine idle.
At $k_b{=}30$ the GPU finishes its prefix $334$~s before the NPU finishes the suffix and the decode.
At $k_b{=}60$ the NPU waits $132$~s for the GPU.
The periods are $12.5\%$ and $14.5\%$ above the deployed cut.
A block boundary narrows that gap to $1$~s with the same single hand-off.

Selecting $k_b$ among block boundaries requires co-run rates and the decoder term $T_V$ of Equation~\ref{eq:kb}.
Balancing block rates alone without $T_V$ gives $k_b{=}34$ and a period $6.3\%$ above the deployed cut because the NPU still owes a decode that no longer fits in the GPU's shadow.
Solo rates give $k_b{=}57$ and a period $10.1\%$ above it because they over-estimate the GPU and hand it $6$ blocks too many.
Co-run rates with $T_V$ give the deployed $k_b{=}51$ and balance the two stages to within $1$~s.
Every period is reported at the deployed run's decoder time $T_V{=}267$~s because that time drifted from $186$ to $270$~s between measurement days.

\begin{table}[t]
\centering
\caption{Where to cut the trajectory, at 49 frames. Each row is the same
pipeline with a different cut index $k_b$, over six videos generated back to back with decoding inside the pipeline. Stage times are per period. Idle is the gap between the two stages.}
\label{tab:cut}
\footnotesize
\setlength{\tabcolsep}{4pt}
\begin{tabular}{@{}llrrrr@{}}
\toprule
Cut selected by & $k_b$ & GPU (s) & NPU (s) & Idle (s) & Period (s) $\downarrow$ \\
\midrule
\multicolumn{6}{@{}l}{\emph{Step boundary: the only positions a step-level partition offers}} \\
step before $k^*$      & 30 & 431 & 765 & 334 & 768 \phantom{0}($+12.5\%$) \\
step after $k^*$       & 60 & 782 & 650 & 132 & 782 \phantom{0}($+14.5\%$) \\
\multicolumn{6}{@{}l}{\emph{Block boundary: the rate model chooses the index}} \\
block rates alone      & 34 & 443 & 723 & 280 & 726 \phantom{00}($+6.3\%$) \\
solo rates             & 57 & 747 & 655 & \phantom{0}92 & 752 \phantom{0}($+10.1\%$) \\
\textbf{co-run rates, with $T_V$} & \textbf{51} & \textbf{676} & \textbf{675} & \textbf{1} & \textbf{683} \\
\bottomrule
\end{tabular}
\end{table}

\begin{figure}[t]
\centering
\includegraphics[width=\columnwidth]{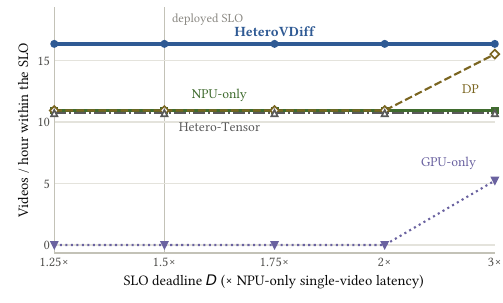}
\caption{Videos per hour within the SLO against the deadline factor $f$ at 17
frames, where the deadline is $D = f\times$ the NPU-only single-video latency.
\sysname clears every deadline in the sweep since its worst video takes
$1.24{\times}$ that latency.
DP and GPU-only both need $2.09{\times}$, and below that DP delivers exactly the
NPU-only rate.
The vertical rule marks the deployed $1.5{\times}$ SLO.}
\Description{Line chart with five lines over deadline factors 1.25 to 3 on the
horizontal axis and videos per hour within the SLO on the vertical axis.
HeteroVDiff is flat at 16.4 across the whole axis and is the highest line
throughout. NPU-only is flat at 10.9 and Hetero-Tensor at 10.7. DP holds 10.9
through factor 2 and rises to 15.5 at 3. GPU-only is zero through factor 2 and
reaches 5.2 at 3. A vertical rule marks the deployed factor of 1.5.}
\label{fig:slo_sensitivity}
\end{figure}

\paragraph{SLO sensitivity.}
Figure~\ref{fig:slo_sensitivity} sweeps the deadline factor $f$ from $1.25$ to $3$ at 17 frames, sets $D = f \times 329.3$~s and plots videos per hour within the SLO for the five configurations.
Two thresholds decide the figure.
The worst \sysname video takes $1.24{\times}$ the NPU-only latency because it crosses both engines, while a DP video on the GPU arm takes $2.09{\times}$.
\sysname therefore clears every deadline in the sweep and holds $16.4$ videos per hour, the highest line throughout.
Up to $f{=}2$ every video DP generates on its GPU arm takes longer than $D$ and misses the deadline.
Only the videos from its NPU arm count, and that arm alone completes $10.9$ videos per hour, the same as NPU-only.
At $f{=}3$ the deadline is long enough for the GPU-arm videos to count, so DP rises to $15.5$.
Even with this loose deadline DP stays below \sysname's $16.4$.
GPU-only starts meeting the deadline at the same point and reaches $5.2$.
Hetero-Tensor stays at $10.7$ at every deadline.
\sysname therefore keeps its gain at every deadline an application would realistically set.
Its only losing case is a deadline below $1.24{\times}$, where every cross-engine video misses.
The cut is chosen when the runner starts, so for such a deadline the runner simply runs NPU-only or a small $k_b$.


\section{Related Work}
\label{sec:related}

\paragraph{Heterogeneous inference on mobile SoCs.}
Concurrent execution across a phone's processors is established for language and perception models~\cite{heteroinfer2025,llmnpu,powerinfer2,nnstretch2023,haxconn,band,pointsplit}, and differs from \sysname{} in two ways.
They synchronise at every layer or tensor slice, about $400$~\textmu s per crossing by HeteroInfer's measurement~\cite{heteroinfer2025}.
Video diffusion makes that far more expensive because every block runs once per denoising step.
Our six-step, 30-block workload executes $N{\cdot}L{=}180$ blocks per video, so a per-layer scheme would cross at least 180 times where \sysname{} crosses once.
They also keep weights in each runtime's own layout. HeteroInfer gives the engines disjoint slices of one matrix. And BlastNet, CoDL, and $\mu$Layer duplicate or reformat blocks per processor~\cite{blastnet,codl,ulayer}.
HeRo aliases one buffer across the QNN and OpenCL runtimes through \texttt{CL\_MEM\_USE\_HOST\_PTR}~\cite{hero} only for inter-stage activations, while weights stay one copy per backend.
None lets two runtimes consume the same persistent weight bytes, the problem §\ref{ssec:design:sharedw} solves.

\paragraph{Cross-engine interference.}
Co-execution on mobile SoCs has been characterised as a cost, or at best as aggregate bandwidth recovered from idle memory channels~\cite{heteroinfer2025}.
HaX-CoNN models GPU--DSP contention as slowdown~\cite{haxconn}, Agent.xpu reports NPU--GPU interference on an AI PC as degradation or neutrality~\cite{agentxpu}, Sereno finds on this SoC that NPU inference degrades foreground rendering~\cite{sereno}, and the one documented cross-component dynamic voltage and frequency scaling (DVFS) coupling in mobile inference is antagonistic~\cite{dvfsgov}.
Their co-runners are separate models or interactive rendering~\cite{edgesysconc,phasematters}; none places a cut from rates that co-execution itself changes, so \sysname{} measures its rates with both engines running, where they move in opposite directions.

\paragraph{On-device diffusion.}
Prior on-device diffusion makes the model smaller and runs it on one engine.
Image systems pair efficient architectures with GPU-aware kernels on a single engine~\cite{snapfusion,mobilediffusion2024}; Speed Is All You Need already keeps a materialised score matrix with a partially fused softmax on Adreno, reserving fused FlashAttention for head dimension 40~\cite{speedisall}, and we re-adopt that structure at head dimension 128, in INT8, and at video sequence length rather than claim it.
Video systems retrain compact models~\cite{snapgenv,tamingdit}, prune tokens or distil steps~\cite{coredit,mobilevd,neodragon}, or page blocks in on demand~\cite{movd}, all orthogonal to \sysname, which keeps the model and changes where it runs.
The closest vendor systems run entirely on the Hexagon NPU and leave the GPU idle~\cite{neodragon,mobilewan}.
MobileWan also splits its block sequence, but within the NPU to fit memory, unloading and reloading at every boundary; S$^2$DiT gives the GPU only the decoder, so its engines run in sequence~\cite{s2dit}; CODA disaggregates video diffusion across two units, but needs near-memory hardware phones lack and overlaps only classifier-free guidance branches our model lacks~\cite{coda}.

\section{Conclusion}
\label{sec:conclusion}

\sysname serves a multi-expert video diffusion model across the NPU and the GPU of one phone from a single stored copy of its weights.
The base weights live in one INT8 artifact laid out as the NPU reads it and consumed by the GPU as integer dot-product operands. The second expert is a rank-128 factor on top of that base. And the denoising trajectory is cut once at a block boundary so that the GPU runs the prefix of the next video while the NPU finishes the current one.
On a Galaxy~S25 this completes a 49-frame video every $677$~s against $917$~s for NPU-only execution, a stream-throughput gain of $1.35\times$ that rises to $1.50\times$ at 17 frames, while staying within $0.2$~pp of FP32 in VBench Quality score and $1.3$~pp in VBench-2.0 Total score.
The general lesson is that two accelerators should share weight bytes rather than arithmetic on a shared-memory SoC, and the cut between them must be placed from co-run rates that include the decoder.

The results come from one SoC generation with a cut selected offline and applied statically.
The cut is a start-up parameter, so a runner could recompute it between videos from the rates it has just observed. The mechanisms themselves depend only on two accelerators sharing one DRAM, and should carry over to other SoCs of that kind.

\bibliographystyle{ACM-Reference-Format}
\bibliography{sample-base}

@inproceedings{ho2020denoising,
  author    = {Ho, Jonathan and Jain, Ajay and Abbeel, Pieter},
  title     = {Denoising Diffusion Probabilistic Models},
  booktitle = {Advances in Neural Information Processing Systems (NeurIPS)},
  volume    = {33},
  pages     = {6840--6851},
  year      = {2020}
}

@article{song2021ddim,
  title={Denoising diffusion implicit models},
  author={Song, Jiaming and Meng, Chenlin and Ermon, Stefano},
  journal={arXiv preprint arXiv:2010.02502},
  year={2020}
}

@inproceedings{peebles2023dit,
  author    = {Peebles, William and Xie, Saining},
  title     = {Scalable Diffusion Models with Transformers},
  booktitle = {IEEE/CVF International Conference on Computer Vision (ICCV)},
  pages     = {4195--4205},
  year      = {2023}
}

@article{wan2025,
  title={Wan: Open and advanced large-scale video generative models},
  author={Wan, Team and Wang, Ang and Ai, Baole and Wen, Bin and Mao, Chaojie and Xie, Chen-Wei and Chen, Di and Yu, Feiwu and Zhao, Haiming and Yang, Jianxiao and others},
  journal={arXiv preprint arXiv:2503.20314},
  year={2025}
}

@inproceedings{dcm2025,
  author    = {Lv, Zhengyao and Si, Chenyang and Pan, Tianlin and Chen, Zhaoxi and Wong, Kwan-Yee K. and Qiao, Yu and Liu, Ziwei},
  title     = {Dual-Expert Consistency Model for Efficient and High-Quality Video Generation},
  booktitle = {Proceedings of the IEEE/CVF International Conference on Computer Vision (ICCV)},
  year      = {2025},
  note      = {arXiv:2506.03123}
}

@article{ediffi,
  title={ediff-i: Text-to-image diffusion models with an ensemble of expert denoisers},
  author={Balaji, Yogesh and Nah, Seungjun and Huang, Xun and Vahdat, Arash and Song, Jiaming and Zhang, Qinsheng and Kreis, Karsten and Aittala, Miika and Aila, Timo and Laine, Samuli and others},
  journal={arXiv preprint arXiv:2211.01324},
  year={2022}
}

@article{timestepmaster,
  title={Timestep master: Asymmetrical mixture of timestep lora experts for versatile and efficient diffusion models in vision},
  author={Zhuang, Shaobin and Guo, Yiwei and Ding, Yanbo and Li, Kunchang and Chen, Xinyuan and Wang, Yaohui and Wang, Fangyikang and Zhang, Ying and Li, Chen and Wang, Yali},
  journal={arXiv preprint arXiv:2503.07416},
  year={2025}
}

@misc{snapdragon8elite,
  author       = {{Qualcomm Technologies, Inc.}},
  title        = {{Snapdragon 8 Elite Mobile Platform Product Brief}},
  year         = {2024},
  howpublished = {Product Brief},
  url          = {https://www.qualcomm.com/smartphones/products/8-series/snapdragon-8-elite-mobile-platform},
  note         = {Accessed: 2026-09-02}
}

@manual{hexagonv79,
  author       = {{Qualcomm Technologies, Inc.}},
  title        = {{Qualcomm Hexagon V79 Programmer's Reference Manual}},
  organization = {Qualcomm Technologies, Inc.},
  number       = {80-N2040-60},
  year         = {2024},
  url          = {https://docs.qualcomm.com/doc/80-N2040-60},
  note         = {Accessed: 2026-09-02}
}

@manual{adreno830,
  author       = {{Qualcomm Technologies, Inc.}},
  title        = {{Adreno 830} {GPU} Architecture and {OpenCL} Programming Reference},
  organization = {Qualcomm Technologies, Inc.},
  year         = {2024},
  note         = {Qualcomm technical reference manual}
}

@misc{qnn-sdk,
  author       = {{Qualcomm Technologies, Inc.}},
  title        = {{Qualcomm AI Engine Direct SDK}},
  howpublished = {\url{https://www.qualcomm.com/developer/software/qualcomm-ai-engine-direct-sdk}},
  note         = {Accessed: 2026-09-02},
  year         = {2024}
}

@manual{qualcomm-fastrpc,
  author       = {{Qualcomm Technologies, Inc.}},
  title        = {{FastRPC}: Remote Procedure Call Framework for {Hexagon DSP}},
  organization = {Qualcomm Technologies, Inc.},
  note         = {Qualcomm Hexagon SDK documentation},
  year         = {2024}
}

@misc{linux-dmabuf,
  author       = {{Linux Kernel Developers}},
  title        = {Buffer Sharing and Synchronization ({dma-buf})},
  howpublished = {\url{https://www.kernel.org/doc/html/latest/driver-api/dma-buf.html}},
  note         = {The Linux Kernel documentation. Accessed: 2026-09-02},
  year         = {2024}
}

@misc{android-ahb,
  author       = {{Android Developers}},
  title        = {Native Hardware Buffer ({AHardwareBuffer})},
  howpublished = {\url{https://developer.android.com/ndk/reference/group/a-hardware-buffer}},
  note         = {Android NDK API Reference. Accessed: 2026-09-02},
  year         = {2024}
}

@inproceedings{nnstretch2023,
  title={Nn-stretch: Automatic neural network branching for parallel inference on heterogeneous multi-processors},
  author={Wei, Jianyu and Cao, Ting and Cao, Shijie and Jiang, Shiqi and Fu, Shaowei and Yang, Mao and Zhang, Yanyong and Liu, Yunxin},
  booktitle={Proceedings of the 21st Annual International Conference on Mobile Systems, Applications and Services},
  pages={70--83},
  year={2023}
}

@inproceedings{heimdall2024,
  title={Heimdall: mobile GPU coordination platform for augmented reality applications},
  author={Yi, Juheon and Lee, Youngki},
  booktitle={Proceedings of the 26th Annual International Conference on Mobile Computing and Networking},
  pages={1--14},
  year={2020}
}

@misc{mlc-llm,
  title={Mlc llm: Universal llm deployment engine with ml compilation},
  author={LLM, MLC},
  year={2025}
}

@article{executorch,
  title={ExecuTorch-A Unified PyTorch Solution to Run ML Models On-Device},
  author={Nachin, Mergen and Desai, Digant and Jia, Stephen and Lai, Chen and Liu, Mengwei and Szwejbka, Jacob and Alvarez, Raziel and Ascani, RJ and Bort, Dave and Candales, Manuel and others},
  journal={Proceedings of Machine Learning and Systems},
  volume={8},
  pages={569--609},
  year={2026}
}

@article{openai2024sora,
  title={Video generation models as world simulators},
  author={Brooks, Tim and Peebles, Bill and Holmes, Connor and DePue, Will and Guo, Yufei and Jing, Li and Schnurr, David and Taylor, Joe and Luhman, Troy and Luhman, Eric and Ng, Clarence and Wang, Ricky and Ramesh, Aditya},
  journal={OpenAI Blog},
  year={2024}
}

@article{hunyuanvideo2024,
  title={Hunyuanvideo: A systematic framework for large video generative models},
  author={Kong, Weijie and Tian, Qi and Zhang, Zijian and Min, Rox and Dai, Zuozhuo and Zhou, Jin and Xiong, Jiangfeng and Li, Xin and Wu, Bo and Zhang, Jianwei and others},
  journal={arXiv preprint arXiv:2412.03603},
  year={2024}
}

@inproceedings{cogvideox2024,
  title={Cogvideox: Text-to-video diffusion models with an expert transformer},
  author={Yang, Zhuoyi and Teng, Jiayan and Zheng, Wendi and Ding, Ming and Huang, Shiyu and Xu, Jiazheng and Yang, Yuanming and Hong, Wenyi and Zhang, Xiaohan and Feng, Guanyu and Yin, Da and Zhang, Yuxuan and Wang, Weihan and Cheng, Yean and Xu, Bin and Gu, Xiaotao and Dong, Yuxiao and Tang, Jie},
  booktitle={International Conference on Learning Representations (ICLR)},
  year={2025},
  note={arXiv:2408.06072}
}

@article{google2024veo,
  title={Video models are zero-shot learners and reasoners},
  author={Wiedemer, Thadd{\"a}us and Li, Yuxuan and Vicol, Paul and Gu, Shixiang Shane and Matarese, Nick and Swersky, Kevin and Kim, Been and Jaini, Priyank and Geirhos, Robert},
  journal={arXiv preprint arXiv:2509.20328},
  year={2025}
}

@article{kapoor2025video_ads,
  title={Frontiers: Generative AI and personalized video advertisements},
  author={Kapoor, Anuj and Kumar, Madhav},
  journal={Marketing Science},
  volume={44},
  number={4},
  pages={733--747},
  year={2025},
  publisher={INFORMS}
}

@inproceedings{emo2024,
  author    = {Tian, Linrui and Wang, Qi and Zhang, Bang and Bo, Liefeng},
  title     = {{EMO}: Emote Portrait Alive---Generating Expressive Portrait Videos with Audio2Video Diffusion Model under Weak Conditions},
  booktitle = {European Conference on Computer Vision (ECCV)},
  year      = {2024}
}

@inproceedings{karras2024fashion_vdm,
  title={Fashion-vdm: Video diffusion model for virtual try-on},
  author={Karras, Johanna and Li, Yingwei and Liu, Nan and Zhu, Luyang and Yoo, Innfarn and Lugmayr, Andreas and Lee, Chris and Kemelmacher-Shlizerman, Ira},
  booktitle={SIGGRAPH Asia 2024 Conference Papers},
  pages={1--11},
  year={2024}
}

@inproceedings{yu2025wonderworld,
  title={Wonderworld: Interactive 3d scene generation from a single image},
  author={Yu, Hong-Xing and Duan, Haoyi and Herrmann, Charles and Freeman, William T and Wu, Jiajun},
  booktitle={2025 IEEE/CVF Conference on Computer Vision and Pattern Recognition (CVPR)},
  pages={5916--5926},
  year={2025},
  organization={IEEE}
}

@misc{gdpr-2016,
  author       = {{European Parliament and Council of the European Union}},
  title        = {Regulation ({EU}) 2016/679 ({General Data Protection Regulation})},
  year         = {2016},
  month        = apr,
  howpublished = {Official Journal of the European Union, L 119, pp. 1--88},
  url          = {https://eur-lex.europa.eu/eli/reg/2016/679/oj},
  note         = {27 April 2016}
}

@misc{ccpa-2018,
  author       = {{State of California}},
  title        = {{California Consumer Privacy Act of 2018} ({CCPA})},
  year         = {2018},
  howpublished = {California Civil Code \S\S~1798.100 et seq.},
  note         = {As amended},
  url          = {https://oag.ca.gov/privacy/ccpa}
}

@article{edge-privacy-survey-2021,
  title={When machine learning meets privacy: A survey and outlook},
  author={Liu, Bo and Ding, Ming and Shaham, Sina and Rahayu, Wenny and Farokhi, Farhad and Lin, Zihuai},
  journal={ACM Computing Surveys (CSUR)},
  volume={54},
  number={2},
  pages={1--36},
  year={2021},
  publisher={ACM New York, NY, USA}
}

@article{wang2020edge,
  author    = {Wang, Xiaofei and Han, Yiwen and Leung, Victor C. M. and Niyato, Dusit and Yan, Xueqiang and Chen, Xu},
  title     = {Convergence of Edge Computing and Deep Learning: A Comprehensive Survey},
  journal   = {{IEEE} Communications Surveys \& Tutorials},
  volume    = {22},
  number    = {2},
  pages     = {869--904},
  year      = {2020}
}

@article{diffserve2025,
  title={Diffserve: Efficiently serving text-to-image diffusion models with query-aware model scaling},
  author={Ahmad, Sohaib and Yang, Qizheng and Wang, Haoliang and Sitaraman, Ramesh K and Guan, Hui},
  journal={Proceedings of Machine Learning and Systems},
  volume={7},
  year={2025}
}

@inproceedings{spotserve2024,
  title={Spotserve: Serving generative large language models on preemptible instances},
  author={Miao, Xupeng and Shi, Chunan and Duan, Jiangfei and Xi, Xiaoli and Lin, Dahua and Cui, Bin and Jia, Zhihao},
  booktitle={Proceedings of the 29th ACM International Conference on Architectural Support for Programming Languages and Operating Systems, Volume 2},
  pages={1112--1127},
  year={2024}
}

@inproceedings{mobilediffusion2024,
  title={Mobilediffusion: Instant text-to-image generation on mobile devices},
  author={Zhao, Yang and Xu, Yanwu and Xiao, Zhisheng and Jia, Haolin and Hou, Tingbo},
  booktitle={European Conference on Computer Vision},
  pages={225--242},
  year={2024},
  organization={Springer}
}

@inproceedings{snapgenv,
  title={Snapgen-v: Generating a five-second video within five seconds on a mobile device},
  author={Wu, Yushu and Zhang, Zhixing and Li, Yanyu and Xu, Yanwu and Kag, Anil and Sui, Yang and Coskun, Huseyin and Ma, Ke and Lebedev, Aleksei and Hu, Ju and others},
  booktitle={2025 IEEE/CVF Conference on Computer Vision and Pattern Recognition (CVPR)},
  pages={2479--2490},
  year={2025},
  organization={IEEE}
}

@inproceedings{hao2026htp,
  title={Scaling llm test-time compute with mobile npu on smartphones},
  author={Hao, Zixu and Wei, Jianyu and Wang, Tuowei and Huang, Minxing and Jiang, Huiqiang and Jiang, Shiqi and Cao, Ting and Ren, Ju},
  booktitle={Proceedings of the 21st European Conference on Computer Systems},
  pages={2157--2172},
  year={2026}
}

@inproceedings{heteroinfer2025,
  title={Characterizing mobile soc for accelerating heterogeneous llm inference},
  author={Chen, Le and Feng, Dahu and Feng, Erhu and Wang, Yingrui and Zhao, Rong and Xia, Yubin and Xu, Pinjie and Chen, Haibo},
  booktitle={Proceedings of the ACM SIGOPS 31st Symposium on Operating Systems Principles},
  pages={359--374},
  year={2025}
}

@article{mobilewan,
  title={MobileWan: Closing the Quality Gap for Mobile Video Diffusion},
  author={Ghafoorian, Mohsen and Korzhenkov, Denis and Karjauv, Adil and Lelekas, Ioannis and Fathima, Noor and Stasis, Spyridon and Ackermann, Hanno and van Breugel, Boris and Nagel, Markus and Porikli, Fatih and others},
  journal={arXiv preprint arXiv:2607.06173},
  year={2026}
}

@inproceedings{neodragon,
  title={Neodragon: Mobile video generation using diffusion transformer},
  author={Karnewar, Animesh and Korzhenkov, Denis and Lelekas, Ioannis and Fathima, Noor and Karjauv, Adil and Ghafoorian, Mohsen and Habibian, Amirhossein},
  booktitle={International Conference on Learning Representations},
  volume={2026},
  pages={39506--39537},
  year={2026}
}

@inproceedings{ptq4dm,
  title={Post-training Quantization on Diffusion Models},
  author={Shang, Yuzhang and Yuan, Zhihang and Xie, Bin and Wu, Bingzhe and Yan, Yan},
  booktitle={IEEE/CVF Conference on Computer Vision and Pattern Recognition (CVPR)},
  year={2023}
}

@inproceedings{qdiffusion,
  title={Q-Diffusion: Quantizing Diffusion Models},
  author={Li, Xiuyu and Liu, Yijiang and Lian, Long and Yang, Huanrui and Dong, Zhen and Kang, Daniel and Zhang, Shanghang and Keutzer, Kurt},
  booktitle={IEEE/CVF International Conference on Computer Vision (ICCV)},
  year={2023}
}

@inproceedings{tdq,
  title={Temporal Dynamic Quantization for Diffusion Models},
  author={So, Junhyuk and Lee, Jungwon and Ahn, Daehyun and Kim, Hyungjun and Park, Eunhyeok},
  booktitle={Advances in Neural Information Processing Systems (NeurIPS)},
  year={2023}
}

@article{viditq,
  title={Vidit-q: Efficient and accurate quantization of diffusion transformers for image and video generation},
  author={Zhao, Tianchen and Fang, Tongcheng and Huang, Haofeng and Liu, Enshu and Wan, Rui and Soedarmadji, Widyadewi and Li, Shiyao and Lin, Zinan and Dai, Guohao and Yan, Shengen and others},
  journal={arXiv preprint arXiv:2406.02540},
  year={2024}
}

@manual{hexagonv81hmx,
  author       = {{Qualcomm Technologies, Inc.}},
  title        = {{Qualcomm Hexagon V81 HMX Programmer's Reference Manual}},
  organization = {Qualcomm Technologies, Inc.},
  number       = {80-N2040-62},
  year         = {2026},
  month        = mar,
  note         = {Rev. AA, March 19, 2026},
  url          = {https://docs.qualcomm.com/doc/80-N2040-62/}
}

@inproceedings{salimans2022progressive,
  author    = {Salimans, Tim and Ho, Jonathan},
  title     = {Progressive Distillation for Fast Sampling of Diffusion Models},
  booktitle = {International Conference on Learning Representations (ICLR)},
  year      = {2022},
  url       = {https://openreview.net/forum?id=TIdIXIpzhoI}
}

@article{song2023consistency,
  title={Consistency models},
  author={Song, Yang and Dhariwal, Prafulla and Chen, Mark and Sutskever, Ilya},
  journal={arXiv preprint arXiv:2303.01469},
  year={2023}
}

@inproceedings{causvid,
  author    = {Yin, Tianwei and Zhang, Qiang and Zhang, Richard and Freeman, William T. and Durand, Fredo and Shechtman, Eli and Huang, Xun},
  title     = {From Slow Bidirectional to Fast Autoregressive Video Diffusion Models},
  booktitle = {{IEEE/CVF} Conference on Computer Vision and Pattern Recognition (CVPR)},
  year      = {2025}
}

@article{t2vturbo,
  title={T2v-turbo: Breaking the quality bottleneck of video consistency model with mixed reward feedback},
  author={Li, Jiachen and Feng, Weixi and Fu, Tsu-Jui and Wang, Xinyi and Basu, Sugato and Chen, Wenhu and Wang, William Y},
  journal={Advances in neural information processing systems},
  volume={37},
  pages={75692--75726},
  year={2024}
}

@inproceedings{ernievilg2,
  title={Ernie-vilg 2.0: Improving text-to-image diffusion model with knowledge-enhanced mixture-of-denoising-experts},
  author={Feng, Zhida and Zhang, Zhenyu and Yu, Xintong and Fang, Yewei and Li, Lanxin and Chen, Xuyi and Lu, Yuxiang and Liu, Jiaxiang and Yin, Weichong and Feng, Shikun and others},
  booktitle={2023 IEEE/CVF Conference on Computer Vision and Pattern Recognition (CVPR)},
  pages={10135--10145},
  year={2023},
  organization={IEEE}
}

@article{raphael,
  title={Raphael: Text-to-image generation via large mixture of diffusion paths},
  author={Xue, Zeyue and Song, Guanglu and Guo, Qiushan and Liu, Boxiao and Zong, Zhuofan and Liu, Yu and Luo, Ping},
  journal={Advances in Neural Information Processing Systems},
  volume={36},
  pages={41693--41706},
  year={2023}
}

@inproceedings{mobilevd,
  author    = {Ben Yahia, Haitam and Korzhenkov, Denis and Lelekas, Ioannis and Ghodrati, Amir and Habibian, Amirhossein},
  title     = {Mobile Video Diffusion},
  booktitle = {Proceedings of the IEEE/CVF International Conference on Computer Vision (ICCV)},
  pages     = {19450--19460},
  year      = {2025},
  doi       = {10.1109/ICCV51701.2025.01808},
  note      = {arXiv:2412.07583}
}

@inproceedings{codl,
  author    = {Jia, Fucheng and Zhang, Deyu and Cao, Ting and Jiang, Shiqi and Liu, Yunxin and Ren, Ju and Zhang, Yaoxue},
  title     = {CoDL: Efficient CPU-GPU Co-execution for Deep Learning Inference on Mobile Devices},
  booktitle = {MobiSys '22: Proceedings of the 20th Annual International Conference on Mobile Systems, Applications and Services},
  doi       = {10.1145/3498361.3538932},
  year      = {2022}
}

@inproceedings{band,
  title={Band: coordinated multi-dnn inference on heterogeneous mobile processors},
  author={Jeong, Joo Seong and Lee, Jingyu and Kim, Donghyun and Jeon, Changmin and Jeong, Changjin and Lee, Youngki and Chun, Byung-Gon},
  booktitle={Proceedings of the 20th Annual International Conference on Mobile Systems, Applications and Services},
  pages={235--247},
  year={2022}
}

@inproceedings{ulayer,
  title={$\mu$layer: Low latency on-device inference using cooperative single-layer acceleration and processor-friendly quantization},
  author={Kim, Youngsok and Kim, Joonsung and Chae, Dongju and Kim, Daehyun and Kim, Jangwoo},
  booktitle={Proceedings of the Fourteenth EuroSys Conference 2019},
  pages={1--15},
  year={2019}
}

@inproceedings{pointsplit,
  author    = {Park, Keondo and Choi, You Rim and Lee, Inhoe and Kim, Hyung-Sin},
  title     = {PointSplit: Towards On-device 3D Object Detection with Heterogeneous Low-power Accelerators},
  booktitle = {IPSN '23: Proceedings of the 22nd International Conference on Information Processing in Sensor Networks},
  doi       = {10.1145/3583120.3587045},
  year      = {2023}
}

@inproceedings{blastnet,
  title={Blastnet: Exploiting duo-blocks for cross-processor real-time dnn inference},
  author={Ling, Neiwen and Huang, Xuan and Zhao, Zhihe and Guan, Nan and Yan, Zhenyu and Xing, Guoliang},
  booktitle={Proceedings of the 20th ACM Conference on Embedded Networked Sensor Systems},
  pages={91--105},
  year={2022}
}

@article{powerinfer2,
  author  = {Xue, Zhenliang and Song, Yixin and Mi, Zeyu and Zheng, Xinrui and Xia, Yubin and Chen, Haibo},
  title   = {PowerInfer-2: Fast Large Language Model Inference on a Smartphone},
  journal = {arXiv preprint arXiv:2406.06282},
  year    = {2024}
}

@inproceedings{llmnpu,
  author    = {Xu, Daliang and Zhang, Hao and Yang, Liming and Liu, Ruiqi and Huang, Gang and Xu, Mengwei and Liu, Xuanzhe},
  title     = {Fast On-device LLM Inference with NPUs},
  booktitle = {Proceedings of the 30th ACM International Conference on Architectural Support for Programming Languages and Operating Systems, Volume 1 (ASPLOS '25)},
  doi       = {10.1145/3669940.3707239},
  year      = {2025}
}

@inproceedings{snapfusion,
  author    = {Li, Yanyu and Wang, Huan and Jin, Qing and Hu, Ju and Chemerys, Pavlo and Fu, Yun and Wang, Yanzhi and Tulyakov, Sergey and Ren, Jian},
  title     = {SnapFusion: Text-to-Image Diffusion Model on Mobile Devices within Two Seconds},
  booktitle = {Advances in Neural Information Processing Systems 36 (NeurIPS 2023), Main Conference Track},
  year      = {2023}
}

@inproceedings{speedisall,
  author    = {Chen, Yu-Hui and Sarokin, Raman and Lee, Juhyun and Tang, Jiuqiang and Chang, Chuo-Ling and Kulik, Andrei and Grundmann, Matthias},
  title     = {Speed Is All You Need: On-Device Acceleration of Large Diffusion Models via GPU-Aware Optimizations},
  booktitle = {Proceedings of the IEEE/CVF Conference on Computer Vision and Pattern Recognition (CVPR) Workshops, 2023, pp. 4651-4655},
  year      = {2023}
}

@article{tamingdit,
  title={Taming diffusion transformer for efficient mobile video generation in seconds},
  author={Wu, Yushu and Li, Yanyu and Kag, Anil and Skorokhodov, Ivan and Menapace, Willi and Ma, Ke and Sahni, Arpit and Hu, Ju and Siarohin, Aliaksandr and Sagar, Dhritiman and others},
  journal={arXiv preprint arXiv:2507.13343},
  year={2025}
}

@inproceedings{movd,
  author    = {Kim, Bosung and Lee, Kyuhwan and Jeong, Isu and Cheon, Jungmin and Lee, Yeojin and Lee, Seulki},
  title     = {Mobile-Oriented Video Diffusion: Enabling Text-to-Video Generation on Mobile Devices Without Retraining, Compression, or Pruning},
  booktitle = {Proceedings of the IEEE/CVF Winter Conference on Applications of Computer Vision (WACV), 2026, pp. 3961-3971},
  doi       = {10.1109/WACV61042.2026.00386},
  year      = {2026}
}

@inproceedings{coredit,
  author    = {Li, Zhuojin and Cheng, Hsin-Pai and Cai, Hong and Han, Shizhong and Porikli, Fatih},
  title     = {CoReDiT: Spatial Coherence-Guided Token Pruning and Reconstruction for Efficient Diffusion Transformers},
  booktitle = {Proceedings of the IEEE/CVF Conference on Computer Vision and Pattern Recognition (CVPR) Workshops, 2026},
  year      = {2026}
}

@article{coda,
  title={CODA: Algorithm-Hardware Co-design for Edge Video Diffusion via NMP-Enabled Compute-Cache Operator Disaggregation},
  author={Zhang, Yuanpeng and Wu, YuXuan and Xiao, Yitong and Xue, Chenhao and Ren, Yi and Li, Cong and Yin, Yihan and Niu, Dimin and Sun, Guangyu},
  journal={arXiv preprint arXiv:2607.14908},
  year={2026}
}

@inproceedings{edgesysconc,
  author    = {Ye, Zilin and Huang, Qianyi},
  title     = {Understanding Concurrent DNN Interference on Mobile SoCs},
  booktitle = {Proceedings of the 24th Annual International Conference on Mobile Systems, Applications and Services Workshops (MobiSys Workshops '26), 2026},
  doi       = {10.1145/3812836.3814775},
  year      = {2026}
}

@article{hero,
  title={HeRo: Adaptive Orchestration of Agentic RAG on Heterogeneous Mobile SoC},
  author={Li, Maoliang and Chen, Jiayu and Zheng, Zihao and Li, Ziqian and Sun, Xinhao and Luo, Guojie and Liu, Chenchen and Chen, Xiang},
  journal={arXiv preprint arXiv:2603.01661},
  year={2026}
}

@inproceedings{haxconn,
  author    = {Dagli, Ismet and Belviranli, Mehmet E.},
  title     = {Shared Memory-contention-aware Concurrent DNN Execution for Diversely Heterogeneous System-on-Chips},
  booktitle = {PPoPP '24: Proceedings of the 29th ACM SIGPLAN Annual Symposium on Principles and Practice of Parallel Programming, 2024},
  doi       = {10.1145/3627535.3638502},
  year      = {2024}
}

@article{agentxpu,
  title={Agent. xpu: Efficient scheduling of agentic llm workloads on heterogeneous soc},
  author={Wei, Xinming and Zhang, Jiahao and Li, Haoran and Chen, Jiayu and Guan, Haoning and Qu, Rui and Li, Maoliang and Chen, Xiang and Luo, Guojie},
  journal={arXiv preprint arXiv:2506.24045},
  year={2025}
}

@article{dvfsgov,
  title={Rethinking DVFS for Mobile LLMs: Unified Energy-Aware Scheduling with CORE},
  author={Zhang, Zongpu and Dash, Pranab and Xu, Qiang and Hu, Y Charlie and Li, Jian and Guan, Haibing},
  journal={Proceedings of Machine Learning and Systems},
  volume={8},
  pages={186--198},
  year={2026}
}

@inproceedings{sereno,
  title={Inference in the Shadows: Taming Memory Bandwidth Contention in Mobile $\{$LLM$\}$ Inference with Sereno},
  author={Xin, Tong and Shi, Xinrui and Dong, Mingkai and Mi, Zeyu},
  booktitle={20th USENIX Symposium on Operating Systems Design and Implementation (OSDI 26)},
  pages={2299--2318},
  year={2026}
}

@inproceedings{phasematters,
  title={Phase Matters: Characterizing Heterogeneous Vision-Language Inference on a Mobile SoC},
  author={Murthy, Aryama V and Kotre, Yashas N and Sharma, Prathmesh and Mishra, Pragya and Ganapathi, Sanjith and Shukla, Priyesh},
  booktitle={Proceedings of the 24th Annual International Conference on Mobile Systems, Applications and Services Workshops},
  pages={322--327},
  year={2026}
}

@article{s2dit,
  title={S2dit: Sandwich diffusion transformer for mobile streaming video generation},
  author={Zhao, Lin and Wu, Yushu and Lebedev, Aleksei and Lahiri, Dishani and Dong, Meng and Sahni, Arpit and Vasilkovsky, Michael and Chen, Hao and Hu, Ju and Siarohin, Aliaksandr and others},
  journal={arXiv preprint arXiv:2601.12719},
  year={2026}
}

@inproceedings{vbench,
  title={Vbench: Comprehensive benchmark suite for video generative models},
  author={Huang, Ziqi and He, Yinan and Yu, Jiashuo and Zhang, Fan and Si, Chenyang and Jiang, Yuming and Zhang, Yuanhan and Wu, Tianxing and Jin, Qingyang and Chanpaisit, Nattapol and others},
  booktitle={2024 IEEE/CVF Conference on Computer Vision and Pattern Recognition (CVPR)},
  pages={21807--21818},
  year={2024},
  organization={IEEE}
}

@article{vbench2,
  title={Vbench-2.0: Advancing video generation benchmark suite for intrinsic faithfulness},
  author={Zheng, Dian and Huang, Ziqi and Liu, Hongbo and Zou, Kai and He, Yinan and Zhang, Fan and Gu, Lulu and Zhang, Yuanhan and He, Jingwen and Zheng, Wei-Shi and others},
  journal={arXiv preprint arXiv:2503.21755},
  year={2025}
}

\appendix
\newpage
\section{Background Details}
\label{app:background}
This appendix keeps the material that §\ref{sec:background} summarises. Reviewers do not need it to follow the paper.

\subsection{Multi-expert models and the DCM sampler}
\label{app:bg:model}
Multi-expert diffusion models specialise different timestep or noise regimes with different denoiser parameters. 
For example, Wan2.2-A14B~\cite{wan2025} stores separate $14$~B high-noise and low-noise transformers and selects between them using a fixed timestep boundary.
TimeStep Master~\cite{timestepmaster} attaches multiple timestep-specialised LoRA experts to a shared backbone, including eight fine-grained core experts and four coarser context experts. 
These experts are lightweight adapters rather than independently stored denoisers.  
DCM~\cite{dcm2025} freezes its distilled semantic expert and uses it as a shared backbone. It forms the detail expert by adding new timestep-embedding layers and rank-128 LoRA adapters to the attention projections. Figure~\ref{fig:dcm} shows its inference pipeline.
Together, the two $14$~B INT8 experts of Wan2.2-A14B require $28$~GB. Even if the pair were factorised post hoc into a shared base plus low-rank adapters, the $14$~B base alone would still exceed our target platform's memory budget unless it were distilled to a smaller model. 
DCM shares parameters by construction. Its semantic expert is the shared base of $1{,}419$~M parameters, and the detail expert adds only a rank-128 attention adapter of $94.4$~M parameters and a new timestep-embedding block of $25.6$~M parameters. 
We therefore use the two expert model DCM-Wan2.1 as our implementation and evaluation workload. This choice is dictated by storage feasibility rather than convenience.

The expert boundary is a timestep threshold, and where it lands in \emph{step index} depends on the sampler. DCM uses the detail expert when $t{\leq}978$. Its sampler uses six fixed timesteps:
\[
[1000,\;992,\;982,\;949,\;905,\;810].
\]
Write $\kappa$ for the index of the first step that uses the detail expert. Here $\kappa{=}3$, so the first three steps use the semantic expert and the final three use the detail expert. 
Two properties of this model matter for the system. 
First, a step is one forward pass and offers no independent second branch to overlap. 
Second, its six steps provide too few partition points to balance the GPU and NPU effectively. 
We therefore consider finer-grained partitioning in §\ref{sec:design}.

\subsection{Snapdragon 8 Elite silicon}
\label{app:bg:silicon}
\begin{figure}[t]
  \centering
  \includegraphics[width=\columnwidth]{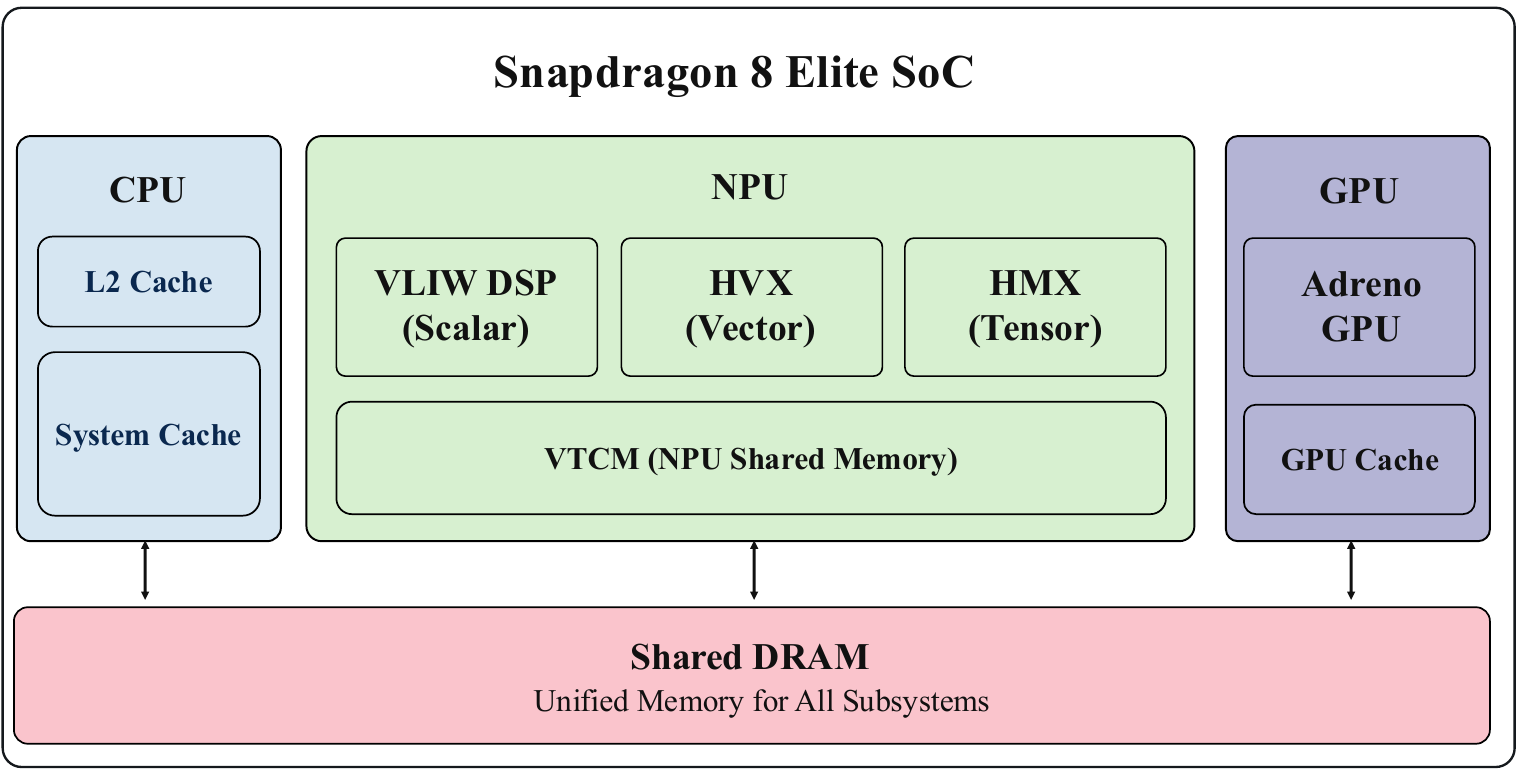}
  \caption{The Snapdragon~8~Elite SoC. The Hexagon NPU pairs a VLIW scalar DSP
  with the HVX vector unit and the HMX matrix unit over VTCM, its
  software-managed on-chip memory. The CPU and the Adreno GPU keep their own
  caches, and all three subsystems reach the same shared DRAM.}
  \Description{Block diagram of the Snapdragon 8 Elite SoC. A CPU block with L2
  and system caches, an NPU block containing a VLIW DSP, an HVX vector unit, an
  HMX tensor unit and VTCM, and a GPU block with the Adreno GPU and its cache.
  All three subsystems connect downward to one shared DRAM.}
  \label{fig:soc}
\end{figure}

The Snapdragon 8 Elite SoC~\cite{snapdragon8elite} used in Samsung Galaxy S25 platform exposes two programmable accelerators that share a unified memory pool, shown in Figure~\ref{fig:soc}. The Hexagon V79 NPU contains a VLIW DSP, a 1024-bit SIMD vector unit (HVX), and a matrix unit (HMX).
HMX multiplies a $32{\times}32$ tile of activations by a $32{\times}32$ tile of weights. 
Both operands must already sit in VTCM, in the tile order the hardware expects~\cite{hexagonv81hmx}. 
HMX supports INT2, INT4, INT8, INT16, FP8, and FP16 with mixed-precision support~\cite{snapdragon8elite}. 
Reachable throughput is gated by the FastRPC~\cite{qualcomm-fastrpc} channel between the application processor and the compute DSP (cDSP), the subsystem that hosts the NPU. 
Programs are typically delivered as offline-compiled Qualcomm AI Engine Direct (QNN) context binaries~\cite{qnn-sdk}. 
The toolchain hardcodes input shapes at build time and refuses to serialise graphs whose activation footprint exceeds an internal limit, a constraint we revisit in §\ref{sec:eval}.

The Adreno 830 GPU~\cite{adreno830} exposes INT8 compute through an OpenCL integer dot-product instruction (\texttt{dp4a}). 
It has no matrix unit. 
HMX keeps a tile resident and reuses it in hardware. 
On the GPU every load from DRAM must be scheduled and reused by the kernel itself. 
A general matrix multiplication (GEMM) reuses each load heavily and attention barely does, which is why the GPU slows down at attention.

The system kernel reports $10.85$~GiB of usable memory on this device.
Adreno advertises a maximum OpenCL buffer size of $1024$~MiB, and \S\ref{sec:motivation} shows that the actual limit is lower.
The two runtimes share memory only through a DMA-BUF~\cite{linux-dmabuf} file descriptor, allocated either through Android's \texttt{AHardwareBuffer} API~\cite{android-ahb} or through the Hexagon runtime's own heap.
One descriptor can be mapped by the CPU, the GPU, and the DSP at the same time.
QNN's Hexagon backend accepts memory only as such a descriptor and has no way to take an OpenCL buffer directly.
A tensor produced by a GPU kernel therefore reaches an NPU graph only when both runtimes register the same region.

\subsection{Heterogeneous inference for generative models}
\label{app:bg:hetero}
Two strategies dominate prior work on splitting an inference workload across multiple accelerators on the same SoC. 
\emph{Layer partition} (\eg HeteroInfer~\cite{heteroinfer2025}, NN-Stretch~\cite{nnstretch2023}, and Heimdall~\cite{heimdall2024}) places a contiguous prefix of model layers on one device and the suffix on another, handing off the activation tensor at each boundary. This strategy incurs at least one cross-device handoff per layer split. An LLM crosses that boundary once per token. A diffusion model would cross it once per block of every step, which is $N{\cdot}L$ times for a single video. 
\emph{Expert partition} is used in mobile MoE deployment~\cite{mlc-llm,executorch}, which places different experts on different devices and routes tokens dynamically.

The denoising trajectory gives diffusion models a third axis that LLMs do not have. 
A video synthesiser runs the same DiT graph $N$ times against a common latent, with no token-level dependency between steps.
Few-step distillation has pushed $N$ from $25{-}50$ down to $4{-}8$, which shrinks the trajectory but does not remove the axis. 
\emph{Step partition} runs steps $0..k{-}1$ on one accelerator and steps $k..N{-}1$ on another, with a single latent handoff in the middle. The handoff cost is then $O(1)$ per video rather than $O(N{\cdot}L)$ for $L$ blocks. 
This matters because every synchronisation between the two accelerators has a fixed cost~\cite{heteroinfer2025}.

\section{Implementation Details}
\label{app:impl}
This appendix expands §\ref{sec:impl} with the kernel-level detail of the NPU and GPU DiT paths and the VAE decoder.

\subsection{NPU DiT Kernels}
\label{ssec:impl:npu}

The NPU executor of Figure~\ref{fig:overview} is a custom FastRPC package. Within it, \sysname ports the open-source \texttt{htp-ops-lib} HMX FlashAttention kernel~\cite{hao2026htp} to the model's video shapes.
The executor creates a six-context HVX worker pool, enables HMX for those workers, and dispatches every attention kernel through that pool.
The deployed workload does not use an attention mask, so the executor takes an unmasked path and never allocates the $M{\times}M$ zero-valued mask tensor that a generic kernel would read.

The port fixes two library routines that were wrong for this model.
The softmax exponential was a lookup table that is accurate only for small attention scores, and at this model's score magnitudes its output correlates with an FP64 reference at only $0.65$.
We replace it with a polynomial \texttt{exp2} in FP16 on HVX, which is correct at every magnitude we tested and costs about $1.2\times$ in attention time.
The FP32-to-FP16 cast, which the executor applies wherever the FP32 residual stream enters a GEMM, used a packing instruction that interleaves its two input vectors, so the outputs came out in the wrong order.
One de-interleave step after it makes the cast bit-exact.

The remaining changes implement the model's attention semantics.
Both self-attention and cross-attention use three-axis rotary positions with adjacent-pair rotation.
The kernels also apply the query-key normalisation gains and all projection
biases supplied by the model.
The executor and kernels share the same rotary tables and metadata.

HMX GEMMs consume FP16 operands, so the NPU dequantises the two feed-forward matrices of the shared INT8 base once per block into a reusable FP16 scratch buffer.
HVX reorders and widens the tiled INT8 values, then applies the per-column
scales.
All output chunks in that block reuse the converted weights before the scratch buffer is recycled for the next block.

For detail-expert blocks, the kernel applies the rank-128 adapter at run time instead of merging it into the weights.
A first HMX GEMM projects the block input onto the 128-dimensional factor basis, a second expands it back, and an HVX pass adds the result into the FP16 projection output.
The base weights never change, so the expert switch moves no weights.

\subsection{GPU DiT Kernels}
\label{ssec:impl:gpu}

The GPU DiT executor of Figure~\ref{fig:overview} implements each block as a sequence of OpenCL kernels that read the shared base-weight artifact directly.
Attention is split into a query-key GEMM, a row-wise softmax, and an attention-value GEMM.
This structure gives the integer dot-product kernels the operand reuse of the GEMM mainloop, and it keeps each dispatch to tens of milliseconds.
Our fused INT8 kernel, which launched the whole sequence as one dispatch, was killed before it completed at the deployment length.

The GPU reads the canonical HMX-tiled INT8 base-weight artifact without creating a reordered copy.
Four fixed-stride vector loads assemble each \texttt{char4} operand, and the per-column weight scale is applied in the GEMM epilogue.
The FP16 inter-block hidden state and the scale for each GEMM input follow §\ref{ssec:design:precision}.


\subsection{NPU VAE Decoder}
\label{ssec:impl:vae}

The NPU runs the VAE decoder through the QNN HTP backend after the final denoising step.
The QNN GPU backend does not provide the required three-dimensional convolution.
On the QNN HTP backend, the floating-point version fails during graph preparation, while the INT8 version compiles.
The decoder is therefore implemented as an INT8 QNN HTP graph.

The first frame and later frames use separate QNN HTP context binaries.
The first frame skips a temporal upsampling operation, while later frames read and update the recurrent feature cache.
A zero-filled cache is not equivalent to the first-frame computation, so the runtime selects the appropriate context for each frame.

The recurrent feature cache also rules out decoding a whole frame at once, because the graph's estimated allocation exceeds the toolchain's 2\,GiB per-graph ceiling.
\sysname therefore divides each latent frame into spatial VAE tiles and uses the largest tile we found to compile under that ceiling, $60{\times}40$ latent pixels, three per frame.
Tiles are decoded one after another, and each keeps its own recurrent cache across all frames of the clip, so only one tile's cache is resident at a time.


\section{VBench-2.0 Scores}
\label{app:vbench2}
Table~\ref{tab:vbench2} lists the per-dimension VBench-2.0 scores discussed in §\ref{ssec:eval:quality}.

\begin{table*}[t]
\centering
\caption{VBench-2.0~\cite{vbench2} scores (\%) of the same four numeric paths on all 18 dimensions, grouped by category. Total is the mean of the five category means.}
\label{tab:vbench2}
\footnotesize
\setlength{\tabcolsep}{1.8pt}
\newcommand{\vbrotb}[1]{\makebox[0pt][c]{\rotatebox[origin=l]{60}{#1}}}
\resizebox{\textwidth}{!}{%
\begin{tabular}{@{}l*{19}{c}@{}}
\toprule
 & \multicolumn{3}{c}{Human fidelity} & \multicolumn{7}{c}{Controllability} & \multicolumn{2}{c}{Creativity} & \multicolumn{4}{c}{Physics} & \multicolumn{2}{c}{Commonsense} & \\
\cmidrule(lr){2-4} \cmidrule(lr){5-11} \cmidrule(lr){12-13} \cmidrule(lr){14-17} \cmidrule(lr){18-19}
Numeric path & \vbrotb{Human anatomy} & \vbrotb{Human identity} & \vbrotb{Human clothes} & \vbrotb{Dynamic attribute} & \vbrotb{Dynamic spatial rel.} & \vbrotb{Motion order} & \vbrotb{Human interaction} & \vbrotb{Complex plot} & \vbrotb{Complex landscape} & \vbrotb{Camera motion} & \vbrotb{Diversity} & \vbrotb{Composition} & \vbrotb{Mechanics} & \vbrotb{Thermotics} & \vbrotb{Material} & \vbrotb{Multi-view consist.} & \vbrotb{Instance preserv.} & \vbrotb{Motion rationality} & \vbrotb{\textbf{Total}} \\
\midrule
FP32 (no quantisation) & 87.93 & 78.07 & 83.03 & 18.32 & 22.55 & 13.85 & 33.33 & 6.41 & 13.11 & 10.19 & 81.33 & 42.06 & 58.73 & 51.45 & 29.11 & 14.49 & 85.38 & 22.41 & 50.77 \\
NPU: INT8 weights, FP16 compute & 88.33 & 76.16 & 83.64 & 22.71 & 22.06 & 16.16 & 31.33 & 7.98 & 12.89 & 11.42 & 81.26 & 41.55 & 63.93 & 53.28 & 30.86 & 9.54 & 84.80 & 22.41 & 50.98 \\
GPU: INT8 weights, INT8 activations & 86.96 & 79.89 & 82.41 & 19.05 & 21.57 & 11.11 & 25.00 & 7.89 & 12.00 & 12.96 & 86.15 & 40.94 & 65.85 & 51.47 & 27.85 & 21.10 & 85.38 & 20.69 & 51.38 \\
\sysname, $k_b{=}51$ & 88.19 & 75.60 & 83.41 & 19.05 & 23.04 & 12.12 & 30.67 & 8.44 & 12.67 & 12.65 & 82.56 & 46.52 & 61.72 & 57.78 & 30.12 & 25.95 & 81.29 & 22.99 & 51.98 \\
\bottomrule
\end{tabular}}
\end{table*}










\end{document}